\documentclass{aa}

\newcommand{\CLASS}{\texttt{CLASS}}

\newcommand{\REPS}{\texttt{REPS}}
\newcommand{\GADGET}[1]{\texttt{GADGET-#1}}

\newcommand{\RAMSES}{\texttt{RAMSES}}

\newcommand{\ANUBIS}{\texttt{ANUBIS}}
\newcommand{\ANUBISIS}{\texttt{ANUBISIS}}

\newcommand{\ISIS}{\texttt{ISIS}}

\newcommand{\rockstar}{\texttt{ROCKSTAR}}
\newcommand{\revolver}{\texttt{Revolver}}
\newcommand{\ngenic}{\texttt{N-GenIC}}

\newcommand{\grafic}{\texttt{Grafic2}}
\newcommand{\corrfunc}{\texttt{Corrfunc}}

\newcommand{\pylians}{\texttt{Pylians}}
\newcommand{\victor}{\texttt{Victor}}
\newcommand{\pycorr}{\texttt{Pycorr}}
\newcommand{\cobaya}{\texttt{Cobaya}}

\newcommand{\euclid}{\textit{Euclid}}

\newcommand{\LCDM}{$\Lambda$CDM}
\newcommand{\fofr}{$f(R)$}
\usepackage{graphicx}
\usepackage{mathrsfs}
\usepackage{float}
\usepackage{txfonts}
\usepackage[hidelinks]{hyperref}
\begin{document}

   \title{The void-galaxy cross-correlation function with massive neutrinos and modified gravity}


   \author{R. Mauland\inst{1}
          \and
          \O. Elgar\o y\inst{1}
          \and
          D. F. Mota\inst{1}
          \and
          H. A. Winther\inst{1}
          }

   \institute{Institute of Theoretical Astrophysics, University of Oslo, P.O.Box 1029 Blindern, N-0315 Oslo, Norway\\
              \email{renate.mauland-hus@astro.uio.no}
             }

   \date{Received 01/03/2023; accepted 03/05/2023}

 
  \abstract{
  Massive neutrinos and \fofr{}-modified gravity have degenerate observational signatures that can impact the interpretation of results in galaxy survey experiments, such as cosmological parameter estimations and gravity model tests. Because of this, it is important to investigate astrophysical observables that can break these degeneracies.  Cosmic voids are sensitive to both massive neutrinos and modifications of gravity and provide a promising ground for disentangling the above-mentioned degeneracies. In order to analyse cosmic voids in the context of non-\LCDM{} (Lambda cold dark matter) cosmologies, we must first understand how well the current theoretical framework operates in these settings. We performed a suite of simulations with the \RAMSES-based {\it N}-body code \ANUBISIS{}, including massive neutrinos and \fofr{}-modified gravity both individually and simultaneously. The data from the simulations were compared to models of the void velocity profile and the void-halo cross-correlation function (CCF). This was done both with the real space simulation data as model input and by applying a reconstruction method to the redshift space data. In addition, we ran Markov chain Monte Carlo (MCMC) fits on the data sets to assess the capability of the models to reproduce the fiducial simulation values of $f\sigma_8(z)$ and the Alcock-Paczy\`{n}ski parameter, $\epsilon$. The void modelling applied performs similarly for all simulated cosmologies, indicating that more accurate models and higher resolution simulations are needed in order to directly observe the effects of massive neutrinos and \fofr{}-modified gravity through studies of the void-galaxy CCF. The MCMC fits show that the choice for the void definition plays an important role in the recovery of the correct cosmological parameters, but otherwise, there is no clear distinction between the ability to reproduce $f\sigma_8$ and $\epsilon$ for the various simulations.}


   \maketitle
%
\section{Introduction}
Cosmic voids are underdense regions in the large-scale structure (LSS) which together with halos, walls, and filaments make up the cosmic web. They have recently become a popular independent probe for cosmological parameters as they provide information about the parameter combination $f\sigma_8(z)$ through the study of redshift space distortions (RSDs) from the imprint left on the quadrupole of the void-galaxy cross-correlation function (CCF) \citep{Cai2016,NadathurPercival2019,Nadathur2019}. Further on, careful modelling of the RSDs around voids originating from the peculiar velocities of galaxies also opens up for the applied fiducial cosmology to be tested through the Alcock-Paczy\`{n}ski effect \citep{AP1979,Sutter2012,Hamaus2022}. In addition to this, the empty nature of cosmic voids makes them sensitive to diffuse components of our Universe, such as massive neutrinos, and effects of modified gravities which scale inversely with density \citep{Zivick2014,Cai2015,Massara2015,Falck2018,Kreisch2019,Fiorini2022}. 

The presence of massive neutrinos suppresses structure formation on scales smaller than the neutrino free-streaming length \citep{Lesgourgues2006}, while \fofr{}-modified gravity enhances it on scales smaller than the Compton wavelength of the scalaron (e.g. \cite{Cataneo2015}). This results in observational degeneracies which have previously been studied in simulations \citep{Baldi2014,Giocoli2018,Contarini2021}. Both massive neutrinos and modified gravity are independently large research fields. Determining the absolute mass scale of neutrinos is not only an important quest within particle physics but also essential in order to understand LSS formation within cosmology. The massive neutrinos make up a fraction of the matter content of the Universe ranging from $0.5-2\%$ based on the lower and upper limits of the sum of the neutrino masses, $0.06 \,\mathrm{eV} \lesssim \sum m_\nu \lesssim 2.4 \,\mathrm{eV}$, provided by particle physics experiments \citep{PDG2022,Aker2022}. This range can be further constrained through cosmological observations. Effects of massive neutrinos on LSS and signatures left in the cosmic microwave background (CMB) \citep{Archidiacono2017} can be used to estimate the sum of the neutrino masses. New space missions, such as \euclid{} \citep{Laureijs2011}, are now sensitive enough to accurately pick up on the effects of massive neutrinos on the matter power spectrum. \euclid{} aims to measure the sum of the neutrino masses to a precision better than $0.03\,\mathrm{eV}$. This will be achieved through a combined analysis of galaxy clustering and weak gravitational lensing. For small neutrino masses, $\sum m_\nu < 0.1\,\mathrm{eV}$, this is precise enough to determine the neutrino mass hierarchy. 

Although constraints from cosmology are typically tighter than the ones obtained from particle physics experiments, they do depend on an assumed cosmological model. Because of this, disentangling degeneracies between massive neutrinos and various cosmologies is important in order to fully determine the neutrino mass scale. Accounting for massive neutrinos when studying alternatives to \LCDM{} (Lambda cold dark matter) is therefore also necessary to make sure that constraints on the model in question are properly estimated. For \fofr{}-modified gravity, the constraints on the model parameters might be slightly alleviated when massive neutrinos are included in the estimates, due to their degeneracy.

The theory of general relativity (GR) has been thoroughly investigated in high density regions, for instance through local solar system tests \citep{Will2014}. To adhere to the resulting measurements, modified gravity theories typically need to be screened as a function of density. This contributes to the constraints on the parameters of these models and also makes deviations from \LCDM{} hard to observe in highly screened environments. Cosmic voids are underdense regions sensitive to both massive neutrinos and \fofr{}-modified gravity. Because of this, voids have been proposed as a means to separate their known degenerate effects. The void size function at high redshifts for large voids is found by \cite{Contarini2021} to be a promising candidate for the task. 

Before further exploring voids in the context of massive neutrinos and \fofr{}-modified gravity, we must first take a look at the theory behind the models developed to extract cosmological information from them. In our case, we investigate the void-galaxy CCF in redshift space, the void velocity profile, and a reconstruction method for putting galaxy positions observed from redshift space back into real space. These are all based on linear theory and, to a degree, a standard \LCDM{} universe. In this paper, we explore the need for change in these models in order to use them to describe voids in a universe with massive neutrinos and \fofr{}-modified gravity. To do so, we compared them to {\it N}-body simulations containing massive neutrinos and \fofr{}-modified gravity, both independently and combined. We studied how well the model for the void velocity profile and the void-halo CCF in redshift space fit the simulation data in each case. The latter was done both by using all the real space information from the simulations directly as model input and also by treating the simulation data as observational data, using redshift space halo positions and a reconstruction method to gain an estimate of the corresponding real space positions of the halos. In addition, we performed Markov chain Monte Carlo (MCMC) fits for $f\sigma_8(z)$ and the Alcock-Paczy\`{n}ski parameter, $\epsilon$, in order to test how well we are able to recover the fiducial cosmology.

This paper is structured as follows: We start by presenting background theory for cosmic voids, massive neutrinos, and \fofr{}-modified gravity in Sect.~\ref{sec:theory} and then explain our simulation set-up in Sect.~\ref{sec:simulations}. In Sect.~\ref{sec:method} we detail the methodical approach of the paper before reporting our results in Sect.~\ref{sec:results}. Finally, we conclude in Sect.~\ref{sec:conclusions}.


\section{Theory}
\label{sec:theory}
\subsection{Cosmic voids}
In the following subsections, we recap some models developed to extract information from cosmic voids. This includes the modelling of the void-galaxy CCF in redshift space, the velocity profile, and a reconstruction process. In addition, we briefly explain how voids can be used to test cosmology through the Alcock-Paczy\'{n}ski effect. The numerical definition of voids in this work is addressed in Sect.~\ref{sec:voidfinder}.

\subsubsection{Cross-correlation function}
\label{sec:ccf_theory}
The void-galaxy CCF gives us information about how galaxies are distributed around voids as a function of the void-centre galaxy separation \citep{Cai2016,NadathurPercival2019, Woodfinden2022}. In real space, we denote this separation vector by $\mathbf{r}$, and in redshift space by $\mathbf{s}$. As we observe in redshift space, deconstructing the separation vectors into components perpendicular and parallel to the line-of-sight (LOS) is advantageous, and gives the following relations
\begin{align}
    s_\perp &= r_\perp,\\
    s_\parallel &= r_\parallel + \frac{v_{\parallel}}{aH}.
\end{align}
Here, $H$ is the Hubble rate, $a$ the scale factor, and $v_{\parallel}$ the component of the galaxy peculiar velocity parallel to the LOS. The LOS can be defined as from the observer to the void centre, the galaxy or the midpoint between them. We use the observer to void-centre definition in this paper.

The void-galaxy CCF in redshift space, $\xi^s(\mathbf{s})$, can be related to the void-galaxy CCF in real space, $\xi^r(\mathbf{r})$, by the streaming model \citep{Peebles1980,Fisher1995,Paillas2021}:
\begin{align}
    1 + \xi^s(\mathbf{s}) = \int(1 + \xi^r(\mathbf{r}))P(v_\parallel,\mathbf{r})\mathrm{d}v_\parallel \label{eq:streaming},
\end{align}
where $P(v_\parallel,\mathbf{r})$ is the probability distribution function (PDF) of the galaxies' peculiar velocities parallel to the LOS. This is a mapping from real to redshift space, based on the fact that void-galaxy pairs are conserved when moving between the two\footnote{In order for this to hold, the voids must be consistently identified either in real or redshift space for both $\xi^r(\mathbf{r})$ and $\xi^s(\mathbf{s})$. However, identification in redshift space leads to an orientation-dependent selection bias of the possible voids \citep{Chuang2017,Nadathur2019recon}. To avoid this, we always identify voids in real space in this paper.}. The galaxy velocities can further be separated into two components, one describing the coherent outflow velocity away from the void centre, and the other an additional stochastic motion of the galaxies. The first term can be assumed spherically symmetric around and radially directed out from the void centre when averaging over a large number of voids, due to statistical isotropy in real space. Following this argument, we can write the LOS galaxy velocity component as
\begin{align}
    v_\parallel(r,\mu_r) = v_r(r)\mu_r + \Tilde{v}_\parallel,
\end{align}
where $\Tilde{v}_\parallel$ is the random velocity component parallel to the LOS, $\mu_r = \cos\theta$, and $\theta$ is the angle between the LOS and $\mathbf{r}$ so that $v_r(r)\mu_r$ is the part of the coherent outflow velocity profile projected along the LOS. 

By performing a coordinate change from $v_\parallel \to \Tilde{v}_\parallel$, as shown in \cite{Woodfinden2022}, Eq.~(\ref{eq:streaming}) can be written as
\begin{align}
    1 + \xi^s(\mathbf{s}) = \int(1 + \xi^r(\mathbf{r}))P(\Tilde{v}_\parallel,\mathbf{r})J_{\mathbf{rs}}\mathrm{d}\Tilde{v}_\parallel \label{eq:streaming_coord_change},
\end{align}
where $J_{\mathbf{rs}}$ is the Jacobian resulting from the shift from real to redshift space and $P(\Tilde{v}_\parallel,\mathbf{r})$ is the PDF of the random velocity component centred around zero. Writing out the Jacobian we get the expression
\begin{align}
    J_{\mathbf{rs}} =\Bigg[1+\frac{v_r}{raH} + \frac{(v'_r - v_r/r)}{aH}\mu_r^2\Bigg]^{-1},
\end{align}
with derivatives with respect to $r$ denoted by a prime. Simulations show that $P(\Tilde{v}_\parallel,\mathbf{r})$ is approximated well by a Gaussian \citep{NadathurPercival2019,Paillas2021} and we therefore consider
\begin{align}
    P(\Tilde{v}_\parallel,\mathbf{r}) = \frac{1}{\sqrt{2\pi\sigma^2_{v_{\parallel}}(r)}}\exp{\Bigg(-\frac{\Tilde{v}^2_\parallel}{2\sigma^2_{v_{\parallel}}(r)}\Bigg)},
\end{align}
with the additional assumption of a spherically symmetric velocity dispersion, $\sigma^2_{v_{\parallel}}(r)$.

\subsubsection{Velocity modelling}
\label{sec:vel_modelling}
Equation~(\ref{eq:streaming_coord_change}) depends on the coherent mean outflow velocity profile around the void, $\mathbf{v}(r) = v_r(r)\hat{\mathbf{r}}$. This quantity can be provided to the model through a template, for example, obtained from a simulation, or it can be modelled separately. Typically, using linear perturbation theory and the continuity equation \citep{Peebles1980,Peebles1993}, the expression 
\begin{align}
    v_r^\mathrm{Lin}(r) = -\frac{1}{3}faH\Delta(r)r\label{eq:vr_lin}
\end{align}
is used for the radial outflow velocity profile. Here, $f=\mathrm{d}\ln D/\mathrm{d}\ln a$ is the linear growth rate, with $D$ as the growth factor, and $\Delta(r)$ is the integrated density,
\begin{align}
    \Delta(r) = \frac{3}{r^3}\int_0^r\delta(x)x^2dx,
\end{align}
with $\delta(r)$ the dark matter overdensity profile around the void. \cite{Nadathur2019} show that $\Delta(r)\propto\sigma_8$, which leads to $v_r(r)$ depending on the combination of the parameters $f\sigma_8$, where $\sigma_8$ is the amplitude of the linear matter power spectrum at a scale of $8\,h^{-1}\mathrm{Mpc}$. It is also worth noting that in the process of deriving Eq.~(\ref{eq:vr_lin}), the growth rate is assumed constant, which does not hold for cosmologies with massive neutrinos or modified gravity, where the growth rate is scale-dependent (e.g. \cite{Hernandez2017,Mirzatuny2019}). 


Although Eq.~(\ref{eq:vr_lin}) is commonly used to model the coherent outflow velocity, the expression in Eq.~(\ref{eq:streaming_coord_change}) should hold for any spherically symmetric velocity profile. One other example of this is the more general velocity profile from \cite{Paillas2021}, which introduces an extra degree of freedom,
\begin{align}
    v_r^\mathrm{Gen}(r) = -\frac{1}{3}\frac{faH\Delta(r)r}{1+A_v\xi_0(r)}. \label{eq:general_vr}
\end{align}
Here, $\xi_0(r)$ is the void-galaxy CCF monopole and $A_v$ is a free parameter. This expression is able to better fit the void velocity profile by adjusting $A_v$. However, the value of $A_v$ must be estimated from, for example, simulations if we want to use Eq.~(\ref{eq:general_vr}) for void-galaxy CCF modelling.

\subsubsection{Reconstruction}
\label{sec:reconstruction}
When working with observational data, we only have access to galaxy positions in redshift space. However, the model of the void-galaxy CCF in redshift space depends on the real space void-galaxy CCF. One way to obtain this quantity is by using a reconstruction method to estimate the positions of galaxies in real space. 

In the context of voids, \cite{Nadathur2019recon} propose one such method, based on \cite{Nusser1994}. It involves solving the Zel'dovich equation \citep{Zeldovich1970}, 
\begin{align}
    \nabla\cdot\mathbf{\Psi} + \frac{f}{b}\nabla\cdot(\mathbf{\Psi}\cdot\hat{\mathbf{r}})\hat{\mathbf{r}} = -\frac{\delta_g}{b}, \label{eq:recon}
\end{align}
for the displacement field, $\mathbf{\Psi}$. Here, $f$ is the growth rate and $b$ is the linear bias relating the galaxy overdensity field, $\delta_g$, to the matter overdensity field, $\delta$, in redshift space. The displacement field, $\mathbf{\Psi}$, relates the Eulerian and Lagrangian positions of the galaxies. A part of this displacement, given by
\begin{align}
    \mathbf{\Psi}_{\mathrm{RSD}} = -f(\mathbf{\Psi}\cdot\hat{\mathbf{r}})\hat{\mathbf{r}},
\end{align}
is due to the linear RSDs resulting from the galaxies' peculiar velocities. By shifting the galaxy positions in redshift space by $-\mathbf{\Psi}_{\mathrm{RSD}}$, we can approximately obtain their real space positions. As seen above, this procedure depends on an input value of the growth rate, $f$, and the linear galaxy bias, $b$, which combined make the reconstruction parameter $\beta = f/b$. Consequentially, the reconstruction method depends on a fiducial cosmological model through the growth rate.

\subsubsection{The Alcock-Paczy\'{n}ski effect}
One way to inspect a chosen cosmology is through the Alcock-Paczy\'{n}ski (AP) effect \citep{AP1979}. This test only involves the geometry of the Universe and is performed by studying the ratio of the observed angular and redshift size of an object with a known shape. An example is cosmic voids, which on average should have a spherical configuration. A correctly chosen fiducial cosmology should, when converting from observed redshifts to physical distances, reproduce the spherical shape of the average void. Deviations from this show up as anisotropies in the void-galaxy CCF in addition to the contribution from peculiar velocities. If we can model the RSDs from velocities accurately enough, the AP effect can be used on voids as a test of the fiducial cosmology.

We parameterise the AP effect through the parameters
\begin{align}
    \alpha_{\parallel} = \frac{D_H(z)}{D_H^{\mathrm{fid}}(z)},\quad 
    \alpha_{\perp} = \frac{D_\mathrm{A}(z)}{D_\mathrm{A}^{\mathrm{fid}}(z)},
\end{align}
which are distance ratios parallel and perpendicular to the LOS between the true cosmology and the fiducial cosmological model\footnote{In our case, we already know that the fiducial cosmology should match the true cosmology as we are performing this test on simulated data. Still, this gives us an opportunity to test how well we reproduce the fiducial cosmology and could also be used to estimate the constraining power of an observed data set of similar resolution.}. Here, $D_H(z) = c/H(z)$, which is the Hubble distance at redshift $z$, and $D_\mathrm{A}(z)$ is the comoving angular diameter distance. 

Taking into account the difference between the true cosmology and the fiducial model, the void-galaxy CCF in redshift space scales like
\begin{align}\label{eq:ap_corr}
    \xi^s(s_{\parallel},s_{\perp}) = \xi^{s,\mathrm{fid}}\Big(\alpha_\parallel s_\parallel^{\mathrm{fid}},\alpha_\perp s_\perp^{\mathrm{fid}}\Big)
\end{align}
due to the void-galaxy pair separation, $\mathbf{s}^\mathrm{fid}$, dependency of the calculation. The two AP parameters can be rewritten into
\begin{align}
    \alpha &= \alpha_\perp^{2/3}\alpha_\parallel^{1/3},\\
    \epsilon &= \frac{\alpha_\perp}{\alpha_\parallel},
\end{align}
where $\alpha$ characterises the volume dilation and $\epsilon$ quantifies anisotropic distortions, which largely affect the quadrupole of the CCF, as \cite{Nadathur2019} demonstrate. When performing the calculation of the CCF model, the input parameters $\xi^r(r)$, $\Delta(r)$, and $\sigma_{v_{\parallel}}$ are scaled by $\alpha$ in order to not use the absolute void size as a standard ruler \citep{Nadathur2019recon}. This leaves only a dependence on $\epsilon$. The underlying assumption behind the AP modelling in Eq.~(\ref{eq:ap_corr}) is that the positions of voids in a fiducial cosmology where the AP parameters are not unity, is (in a statistical sense) just an AP stretching of the void positions in the true cosmology. This is not always the case, and the implications of this assumption for extracting constraints from the void-galaxy CCF in future surveys is the topic of an upcoming paper.

\subsection{Massive neutrinos}
From neutrino flavour oscillation experiments, we know that at least two of the three neutrino mass states have a non-zero mass \citep{Fukuda1998,Ahmad2002,Araki2005}. Currently, the flavour oscillation experiments of solar and atmospheric neutrinos put constraints on the differences between the neutrino mass states given by \citep{PDG2022}
\begin{align}
    \Delta m_{21}^2 &= (7.53\pm0.18)\times10^{-5}\,\mathrm{eV}^2,\\
    \Delta m_{32}^2 &= (-2.536\pm 0.034)\times10^{-3}\,\mathrm{eV}^2\;\mathrm{(IH)},\\
    \Delta m_{32}^2 &= (2.453\pm 0.033)\times10^{-3}\,\mathrm{eV}^2\;\mathrm{(NH)}.\
\end{align}
These experiments are not yet sensitive enough to distinguish between the ordering of the mass states, and we, therefore, have two possibilities: the normal hierarchy (NH), where $m_1<m_2\ll m_3$, and the inverted hierarchy (IH), where $m_3\ll m_1 < m_2$. Put together, these measurements provide a lower bound on the sum of the neutrino masses at $\sum m_{\nu}\gtrsim 0.06\,\mathrm{eV}$ and $\sum m_{\nu}\gtrsim 0.1\,\mathrm{eV}$ for the NH and IH respectively. An upper bound on the sum of the neutrino masses is also provided by particle physics and the most recent constraint, at $\sum m_{\nu} \lesssim 2.4\,\mathrm{eV}$, comes from the KATRIN experiment, which investigates the single $\beta$-decay of molecular tritium \citep{Aker2022}. 

Massive neutrinos make up a fraction of the matter content of the Universe, $\Omega_\nu$, described through the relation \citep{Lesgourgues2006}
\begin{align}
\Omega_\nu h^2 \approx \frac{\sum m_\nu}{93.14\,\mathrm{eV}}.
\end{align}
Typically, this is taken out of the dark matter budget, $\Omega_c$, so that the total matter content, $\Omega_m = \Omega_c + \Omega_\nu + \Omega_b$, stays constant. Here, $\Omega_b$ and $\Omega_m$ denote the energy density of baryons and the total matter respectively. As the massive neutrinos are relativistic at early times, they free-stream out of local peaks in the density throughout the first stages of structure formation \citep{Lesgourgues2006}. This, in addition to massive neutrinos altering the background evolution, leads to a suppression of matter fluctuations on scales smaller than the neutrino free-streaming length. The strength of this scale-dependent suppression depends on the neutrino mass and shows up in the linear matter power spectrum at small scales as
\begin{align}
    \frac{\Delta P_{m}}{P_{m}} \approx -8f_\nu.
\end{align}
This is a comparison between a cosmology with massless neutrinos and massive neutrinos, where $P_{m}$ is the linear matter power spectrum considering the total matter and $f_\nu = \Omega_\nu/\Omega_m$. At non-linear scales, this suppression is known to be even stronger, with a maximum around $k\sim 1 \,h\,\mathrm{Mpc}^{-1}$, followed by a turn-around which results in a spoon-like shape of the matter power spectrum ratio \citep{Hannestad2020}. 

The presence of massive neutrinos results in observable features in the LSS and CMB \citep{Will2014}, which makes it possible to obtain an upper bound on the sum of the neutrino masses through cosmological observations. Recently, \cite{Valentino2021} find $\sum m_\nu \lesssim 0.09\,\mathrm{eV}$ at a $95\%$ confidence level by analysing a combination of data sets. This puts pressure on the IH, which might further be confirmed by the \euclid{} space mission \citep{Laureijs2011}. It should, however, be noted that although cosmology can provide a tighter upper bound on the sum of the neutrinos masses than current particle physics experiments, the analysis depends on the choice of a cosmological model.

\subsection{$f(R)$-modified gravity}
\label{sec:fofr}
In the concordance \LCDM{} model, we have the general theory of relativity (GR) with the Einstein-Hilbert action 
\begin{align}
    S = \int \frac{R}{16\pi G} \sqrt{-g}\,\mathrm{d}^4x,
\end{align}
where $R$ is the Ricci scalar and $g$ is the determinant of the metric tensor, $g_{\mu\nu}$. To derive the Einstein field equations, we also need to insert the Lagrangian density of matter, $\mathscr{L}_m$, which describes the matter in the theory,
\begin{align}
    S = \int \left(\frac{R}{16\pi G} + \mathscr{L}_m\right) \sqrt{-g}\,\mathrm{d}^4x.
\end{align}

In \fofr-modified gravity theories, the Einstein-Hilbert action is modified by adding a function depending on the Ricci scalar,
\begin{align}
    S = \int \left(\frac{R + f(R)}{16\pi G} + \mathscr{L}_m\right) \sqrt{-g}\,\mathrm{d}^4x.
\end{align}
For the specific case of Hu-Sawicki \fofr{}-modified gravity, we have
\begin{align}
    f(R)=-m^2\frac{c_1(R/m^2)^n}{c_2(R/m^2)^n+1},
\end{align}
where $m^2=H_0^2\Omega_m$ and $c_1$, $c_2$, and $n$ are constant, dimensionless, non-negative parameters that describe the model \citep{Hu2007}. Requiring that the model gives dark energy in the form of a cosmological constant, these three parameters can be reduced to two, $n$ and $f_{R0}$, where 
\begin{align}
    f_{R0}=-n\frac{c_1}{c_2^2}\left(\frac{\Omega_m}{3(\Omega_m+4\Omega_{\Lambda})}\right)^{n+1} \label{eq:fofr_n}
\end{align}
and $c_1/c_2 = 6\Omega_\Lambda/\Omega_m$. Increasing the value of $|f_{R0}|$ gives greater deviations from GR, which has an effect on structure formation. This was demonstrated in \cite{Hu2007}, where raising the value of $|f_{R0}|$ showed an enhancement in the matter power spectrum on small scales. 

The \fofr{} model has become the `fiducial' model used to investigate cosmological signatures of modified gravity. It is well studied in the literature and constraints have already been produced based on various cosmological probes. For example, an analysis by \cite{Cataneo2015} including cluster, CMB, supernova, and baryon acoustic oscillation (BAO) data obtains an upper bound on the Hu-Sawicki \fofr{} theory given by $\log_{10}|f_{R0}|<-4.79$ for $n=1$ at $95.4\%$ confidence level. This familiarity is the reason why we chose Hu-Sawicki \fofr{} as our modified gravity model for this work. We set the fiducial value to $|f_{R0}|=10^{-5}$, as this is the value that the LSS is currently able to probe \citep{Koyama2016}. In addition, we occasionally include some analysis with $|f_{R0}|=10^{-4}$ to enhance the effects of the modified theory, even though this value has been ruled out by some probes (e.g. \cite{Cataneo2015}).

\subsection{Massive neutrinos and $f(R)$ gravity degeneracy}
Modifications of gravity are often described as adding a fifth force which, in addition to gravity, works attractively. This fifth force is in \fofr{}-modified gravity carried by the scalaron, the scalar degree of freedom, quantified by $\mathrm{d}f/\mathrm{d}R$. The Compton wavelength of the scalaron, $\lambda_C$, determines the range of the fifth force, which also establishes the scale on which Hu-Sawicki \fofr{}-modified gravity enhances structure formation (e.g. \cite{Llinares2014,Cataneo2015}),
\begin{align}
    \lambda_C^0 = 3\sqrt{\frac{(n+1)}{\Omega_m + 4\Omega_\Lambda}}\sqrt{\frac{|f_{R0}|}{10^{-6}}}\,h^{-1}\mathrm{Mpc}.
\end{align} 
For $|f_{R0}|=10^{-5}$ and $n=1$, this corresponds to scales $k \gtrsim 0.1\,h\,\mathrm{Mpc}^{-1}$ at redshift zero. 

Massive neutrinos change the time of matter-radiation equality, leading to changes in the LSS. In addition, they suppress structure formation on scales smaller than the neutrino free-streaming length \citep{Lesgourgues2006},
\begin{align}
    \lambda_\mathrm{FS} = 7.7\frac{1+z}{\sqrt{\Omega_\Lambda + \Omega_m(1+z)^3}}\Bigg(\frac{1\,\mathrm{eV}}{\sum m_\nu}\Bigg)\,h^{-1}\mathrm{Mpc},
\end{align}
due to their inability to cluster as relativistic particles. This corresponds to scales of $k \gtrsim 0.02\,h\,\mathrm{Mpc}^{-1}$ for $\sum m_\nu = 0.15\,\mathrm{eV}$ at redshift zero.

The scales where structure formation is enhanced by \fofr{} gravity coincide with the neutrino free-streaming length. The suppression effect of the massive neutrinos can therefore, depending on neutrino mass and the value of $f_{R0}$, counteract the structure enhancement of \fofr{}-modified gravity. This degeneracy has been shown to affect observables such as the matter power spectrum, the halo mass function (HMF) and the halo bias \citep{Baldi2014}.

\section{Simulations}
\label{sec:simulations}
A variety of simulations incorporating massive neutrinos and modified gravity in different ways already exist (e.g. \cite{MNCCP, Winther2015}), although fewer take both into account simultaneously (e.g. \cite{Baldi2014,Giocoli2018}). We approach this in our own way, by developing a \RAMSES{}-based code, \ANUBISIS{}, which offers the option of both massive neutrinos and several different modified gravity theories. This code is a merger between \ANUBIS{}\footnote{\url{https://github.com/renmau/anubis}} and \ISIS{}, as explained in further detail below.

\subsection{ANUBIS}
\label{sec:anubis}
For massive neutrino simulations, we have implemented neutrino particles in the {\it N}-body and hydrodynamical code \RAMSES{} \citep{Teyssier2002}. \RAMSES{} implements the particle mesh (PM) method with adaptive mesh refinement (AMR), which enables higher resolution in denser regions of the simulation box. As \RAMSES{} is a Newtonian {\it N}-body code, the most straightforward way to add neutrinos is to implement them as their own particle family and alter the equations of motion (EOM) to handle arbitrarily high momenta. Originally, the gravitational potential is calculated by solving the Poisson equation, but we instead solved the geodesic equation written in terms of canonical momentum, as commonly done for {\it N}-body codes including relativistic particles \citep{Ma1994,Adamek2017,MNCCP}. As a first approach, we ignored frame-dragging, scattering of particles on gravitational waves, and anisotropic stress, leaving us with the EOMs
\begin{align}
    \frac{\partial q_i}{\partial \tau} &= -\frac{2q^2 + m^2a^2}{\sqrt{q^2 + m^2a^2}}\frac{\partial\phi_{i}}{\partial x^i},\\
    \frac{\partial x_i}{\partial\tau}&=\frac{q_i}{\sqrt{q^2+m^2a^2}},
\end{align}
where $q$ is the canonical momentum, $a$ is the scale factor, $m$ is the cold dark matter (CDM) particle mass, $\partial\phi/\partial x^i$ is the gravitational force (which for modified gravity models includes additional fifth-forces), $\tau$ is conformal time, and $x_i$ is the coordinate three-position vector.

\begin{figure}
    \centering
    \includegraphics{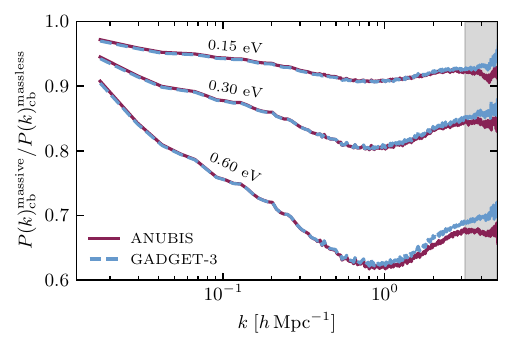}\\
    \vspace{-0.3cm}
    \hspace{-0.2cm}
    \includegraphics{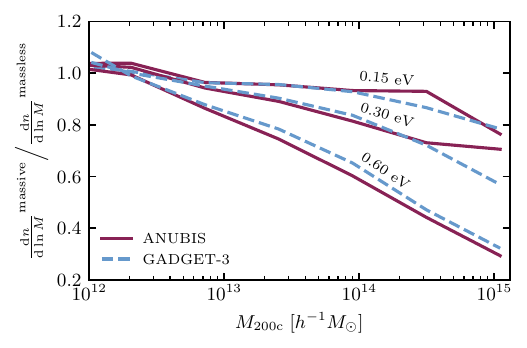}
    \caption{Comparison of the neutrino suppression on the cb power spectrum and HMF for \ANUBIS{} and \GADGET{3} at $z=0$ for various values of the sum of the neutrino masses. \textit{Top}: The neutrino suppression on the cb power spectrum. The greyed-out area indicates the parts of the ratios lying outside of the Nyquist frequency of the simulations. \textit{Bottom}: The neutrino suppression on the HMF. Only halos with $N_{\mathrm{part}}\geq11$ are included.}
    \label{fig:nu_pk_hmf_suppress_anubis_vs_gadget}
\end{figure}

\begin{table*}
\caption{Simulation overview showing the type of simulation and the corresponding properties. The top six entries show the simulations run with the \ANUBISIS{} code, and the bottom two show simulations previously run with the \ISIS{} code.
For the simulations involving Hu-Sawicki \fofr{} gravity, we always set $n=1$ in Eq.~(\ref{eq:fofr_n}). 
}           
\label{table:sim_overview}      
\centering                          
\begin{tabular}{c@{\hskip 0.31in} c@{\hskip 0.31in} c @{\hskip 0.31in}c@{\hskip 0.31in} c@{\hskip 0.31in} c@{\hskip 0.31in} c}        
\hline\hline                 
Simulation &Type&$L_{\mathrm{box}}$ [$h^{-1}\rm{Mpc}$]& $N_{\mathrm{cdm}}$ & $N_{\nu}$ & $|f_{R0}|$&Mass resolution [$h^{-1}M_{\odot}$] \\
\hline                        
   \texttt{lcdm}& $\Lambda$CDM & 1536&$1024^3$ & $0$ & $0$ & $2.99\times10^{11}$ \\
   \texttt{fofr}& $f(R)$ &1536& $1024^3$ & $0$  & $10^{-5}$ & $2.99\times10^{11}$ \\
   \texttt{015ev}& $0.15\;\rm{eV}$ &1536& $1024^3$ &$2\times1024^3$  & $0$ & $2.95\times10^{11}$ \\
   \texttt{06ev}& $0.60\;\rm{eV}$ &1536& $1024^3$ &$2\times1024^3$  & $0$ & $2.85\times10^{11}$ \\
   \texttt{fofr\_015ev}& $f(R)+0.15\;\rm{eV}$  &1536& $1024^3$ &$2\times1024^3$ & $10^{-5}$& $2.95\times10^{11}$ \\
   \texttt{fofr\_06ev}& $f(R)+0.60\;\rm{eV}$  &1536& $1024^3$ &$2\times1024^3$  & $10^{-5}$& $2.85\times10^{11}$ \\
   \hline
   \texttt{lcdm\_small}& \LCDM  &512& $512^3$ & 0  & 0& $8.85\times10^{10}$ \\
   \texttt{fofr\_small}& $f(R)$  &512& $512^3$ &0  & $10^{-4}$& $8.85\times10^{10}$ \\
\hline                                   
\end{tabular}
\end{table*}

In addition to including the massive neutrinos as a new particle family, we also incorporated radiation into the background evolution through the Hubble function, with an additional option to read the Hubble function, $H(z)$, from file. As of now, we have made no changes to how the time-stepping scheme already incorporated in \RAMSES{} operates, which makes for quite slow calculations for the lighter neutrino masses. This could be implemented at a later stage by, for example, detaching the neutrino particles from the grid and allowing them to travel further, letting the CDM particles solely determine the size of the time steps, or by requiring that the neutrino runs use the same refinements and level structure as an equivalent massless neutrino simulation. The latter alternative would allow a more direct comparison between \LCDM{} and massive neutrino simulations, especially when studying various ratio properties.

The performance of the \ANUBIS{} code is tested against other codes incorporating massive neutrinos in the massive neutrino code comparison project (MNCCP) within \textit{Euclid} \citep{MNCCP}, although with a somewhat lower resolution than what is expected for convergence within $1\%$ between the \RAMSES{} and \GADGET{3} codes \citep{Schneider2016}. The upper panel of Fig.~\ref{fig:nu_pk_hmf_suppress_anubis_vs_gadget} displays the neutrino suppression on the CDM + baryon (cb) matter power spectrum for different values of the sum of the neutrino masses at $z=0$ for both \ANUBIS{} and \GADGET{3}. The plot shows good agreement between the two codes, especially for lower neutrino masses. At higher masses, the codes start to deviate at smaller scales. This is most likely due to internal differences in resolution for \RAMSES{}, between the \LCDM{} and massive neutrino simulations. As the neutrino mass increases, structure formation is more suppressed, leading to fewer refinements created by the \RAMSES{} AMR-scheme. This results in a slightly lower resolution for the neutrino simulations, compared to the \LCDM{} one, which shows up in the ratio. This can be resolved by a higher particle density. The lower panel of Fig.~\ref{fig:nu_pk_hmf_suppress_anubis_vs_gadget} displays the neutrino suppression on the HMF for the same neutrino masses as for the matter power spectrum ratios in the upper panel. The magnitude of the suppression is in agreement for both \GADGET{3} and \ANUBIS{}, and increases with the halo mass, corresponding to the findings of \cite{Brandbyge2010}.

\subsection{ISIS}
\label{sec:isis}
\ISIS{} is a cosmological {\it N}-body code incorporating scalar-tensor gravitational theories including screening mechanisms into \RAMSES{} \citep{Llinares2014}. This is done by implementing a non-linear implicit solver for a generic scalar field, which can treat various scalar-tensor-modified gravity theories. This also includes \fofr{} theories, which can be rewritten into the scalar-tensor format. In particular, \ISIS{} contains Hu-Sawicki \fofr{}-modified gravity, as described in Sect.~\ref{sec:fofr}.

\begin{figure}
    \centering
    \includegraphics{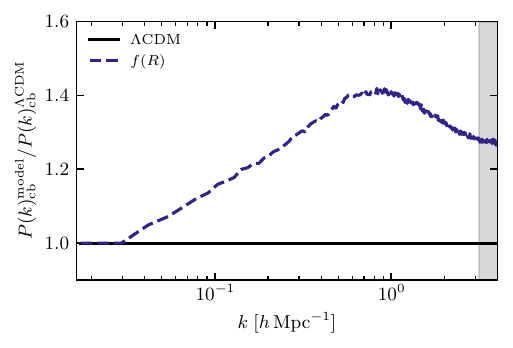}\\
    \vspace{-0.5cm}
    \hspace{-0.2cm}
    \includegraphics{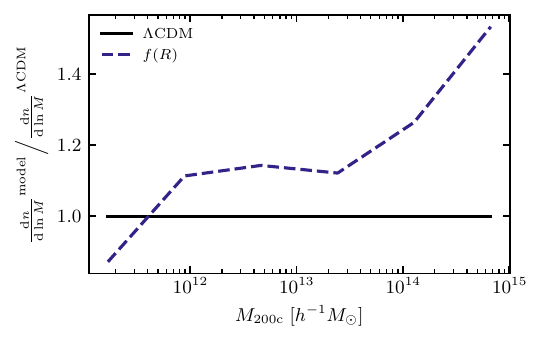}
    \caption{CDM+baryon (cb) matter power spectrum and HMF of the \texttt{fofr\_small} simulation, compared to \texttt{lcdm\_small}. The power spectrum and HMF ratios are displayed in the top and bottom panels respectively. For the power spectrum ratio, the different $\sigma_8$-values of the simulations are taken into account through scaling, and the greyed-out area indicates the parts of the ratios lying outside of the Nyquist frequency of the simulations.}
    \label{fig:f4_pk_ratio_own_sims}
\end{figure}

The upper panel of Fig.~\ref{fig:f4_pk_ratio_own_sims} shows the ratio of the cb matter power spectrum for a \LCDM{} (\texttt{lcdm\_small}) and \fofr{} (\texttt{fofr\_small}) simulation run with \ISIS{}, with the parameter $|f_{R0}|=10^{-4}$. Here, we can clearly see the enhancement of structure at small scales, with a peak at similar scales to the trough in the neutrino suppression in Fig.~\ref{fig:nu_pk_hmf_suppress_anubis_vs_gadget}. For the HMF in the bottom panel of Fig.~\ref{fig:f4_pk_ratio_own_sims}, the amount of large mass halos is increased, compared to \LCDM{}, as opposed to the \ANUBIS{} simulations with massive neutrinos that show a decrease. This enhancement is due to a boost of gravity resulting from the fifth force present in \fofr{}-modified gravity. Summed up, these figures show the opposite behaviour of what is observed in Sect.~\ref{sec:anubis}, as expected.

The two simulations introduced here, dubbed \texttt{lcdm\_small} and \texttt{fofr\_small}, are occasionally used as additions to the simulations performed for this work. This is further explained in the following section, along with the simulation details. 

\begin{figure}
    \centering
    \includegraphics{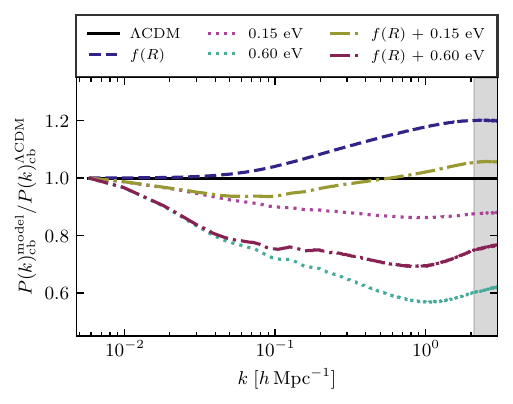}\\
    \hspace{-0.4cm}
    \includegraphics[trim={0 0 0 1.29cm},clip]{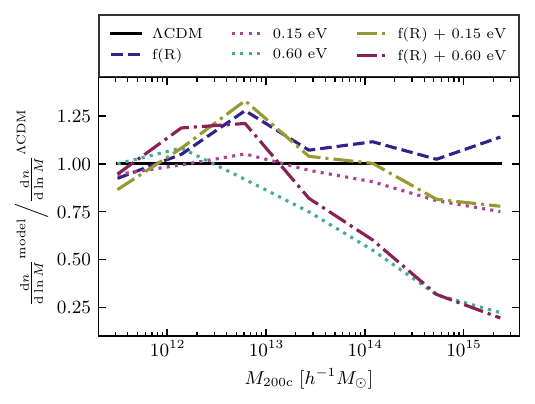}
    \caption{CDM+baryon (cb) matter power spectrum and HMF of the different \ANUBISIS{} simulations, compared to $\Lambda$CDM. The power spectrum and HMF ratios are displayed in the top and bottom panels respectively. For the power spectrum, the greyed-out area indicates the parts of the ratios lying outside of the Nyquist frequency of the simulations.
    }
    \label{fig:pk_HMF_ratio_own_sims}
\end{figure}

\subsection{ANUBIS + ISIS = ANUBISIS}
To obtain a \RAMSES{}-based code which includes both the effects of massive neutrinos and \fofr{} gravity, we have merged the \ANUBIS{} and \ISIS{} codes into one code, \ANUBISIS{}. This provides us with the opportunity to run simulations with massive neutrinos and modified gravity both independently and simultaneously. For this paper, we have performed a suite of such simulations with properties as presented in Table~\ref{table:sim_overview}. Summarised, we have six simulations, one with \LCDM{} cosmology, one with Hu-Sawicki \fofr{}-modified gravity, two with \LCDM{} and massive neutrinos, and two with massive neutrinos and Hu-Sawicki \fofr{} gravity combined. In addition to these six simulations run with the \ANUBISIS{} code, we also have two simulations, one \LCDM{} and one \fofr{}, which was previously run with the \ISIS{} code, as presented in Sect.~\ref{sec:isis}. These are included as the \fofr{} simulation was run with $|f_{R0}|=10^{-4}$, and therefore better demonstrates the effects of modified gravity in specific cases where we are interested in studying this further.

In this section, we present some general properties of the \ANUBISIS{} simulations. To better help distinguish the results, we introduce a general linestyle guide where pure \LCDM{} is always shown as a full line, \fofr{} as a dashed line, massive neutrino runs as a dotted line, and the combination of \fofr{}-modified gravity and massive neutrinos as a dash-dotted line. The upper panel of Fig.~\ref{fig:pk_HMF_ratio_own_sims} shows the ratio of the CDM+baryon (cb) matter power spectrum between the various \ANUBISIS{} simulations and the \LCDM{} case. At large scales, there is originally a slight excess in the power spectrum for the massive neutrino and \fofr{} + massive neutrino runs. This appears as a result of the grid settings used in the simulations, which is further explained in Appendix~\ref{sec:anubisis_res}. Essentially, it amounts to a $\sigma_8$-scaling ($2-3\%$ difference), which is accounted for in the figure. On smaller scales, we observe the expected suppression of structure for the massive neutrino simulations and an enhancement of structure for the \fofr{} simulation. The \fofr{} $+\,0.15\,\mathrm{eV}$ simulation is slightly suppressed, compared to the pure \fofr{} run, but it is still dominated by the modified gravity effects due to the low neutrino mass. The \fofr{} $+\,0.6\,\mathrm{eV}$ simulation, on the other hand, is mostly dominated by the massive neutrinos due to the high neutrino mass, but it is still slightly enhanced, compared to the pure $0.6\,\mathrm{eV}$ run.

\begin{figure}
    \centering
    \includegraphics{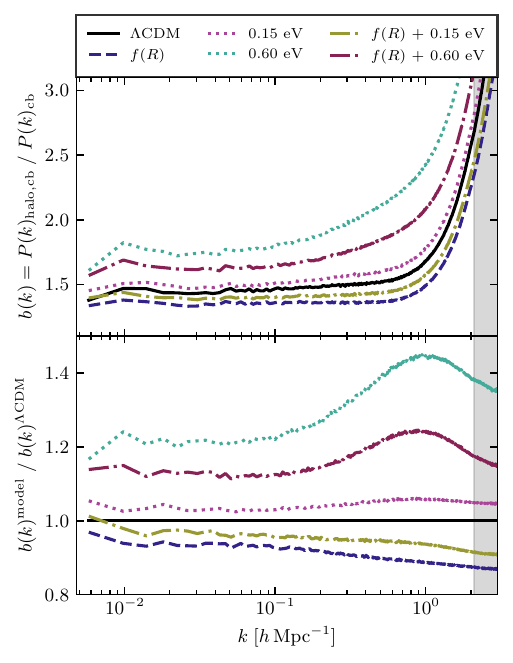}
    \caption{Scale-dependent halo bias for the various \ANUBISIS{} simulations. The upper panel shows the value of the bias for each simulation, while the lower panel shows the ratio with $\Lambda$CDM for each model. The greyed-out area indicates the parts lying outside of the Nyquist frequency of the simulations.}
    \label{fig:halo_bias_own_sims}
\end{figure}

In the bottom panel of Fig.~\ref{fig:pk_HMF_ratio_own_sims}, the HMF of the various simulations, compared to \LCDM{}, is shown. In general, we again see that the massive neutrinos suppress the formation of the most massive halos. This is mostly due to the massive neutrinos reducing the maximum cluster mass on a linear level \citep{Brandbyge2010}. Because of the fifth force contribution to gravity, we again expect the opposite effect for the \fofr{} simulation. This is, however, not very clear for the $|f_{R0}|=10^{-5}$ simulation. If we instead look back at the smaller box simulations from \ISIS{} in Fig.~\ref{fig:f4_pk_ratio_own_sims} with $|f_{R0}|=10^{-4}$, this effect is much more prominent.

\begin{figure}
    \hspace{-0.2cm}
    \includegraphics{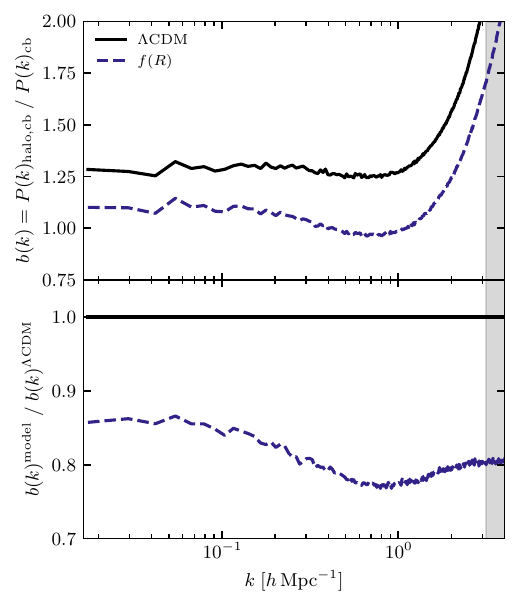}
    \caption{Scale-dependent halo bias for the \texttt{fofr\_small} and \texttt{lcdm\_small} simulations. The upper panel shows the value of the bias for each simulation, while the lower panel shows the ratio with $\Lambda$CDM. The greyed-out area indicates the parts lying outside of the Nyquist frequency of the simulations.}
    \label{fig:f4_halo_bias_own_sims}
\end{figure}

Another interesting property is the scale-dependent halo bias. Here, we define it as the ratio between the cross-power spectrum of halos and CDM+baryons and the auto matter power spectrum of CDM+baryons,
\begin{align}
b(k) = \frac{P(k)_{\mathrm{halo,cb}}}{P(k)_{\mathrm{cb}}}.   
\end{align}
This is defined by the cold species instead of the total matter, as it yields a more universal and scale-independent result in the presence of massive neutrinos \citep{Castorina2014}. In Fig.~\ref{fig:halo_bias_own_sims}, we see the halo bias for the \ANUBISIS{} simulations at the top, and the ratio of the biases, compared to \LCDM{}, at the bottom. The ratios show a bump at the same scales where we see a trough for the massive neutrino power spectrum ratios. In our case, this shows that our halo power spectrum is less sensitive to the neutrino mass than the cb power spectrum, as discussed by \cite{Hassani2022}. In Fig.~\ref{fig:f4_halo_bias_own_sims} we also show the scale-dependent bias for the pure \ISIS{} simulations. In general, we see that the halo bias increases with neutrino mass and decreases with the $f_{R0}$ parameter, compared to \LCDM{}, in line with previous findings in the literature (e.g. \cite{Arnold2019} and \cite{Chiang2019}).

\begin{table}
\caption{Linear bias, reconstruction parameter, and $\sigma_8$ for the various simulations. The linear halo bias, $b$, is found by taking the average of the scale-dependent bias for $k<0.05\;h^{-1}\mathrm{Mpc}$. Values of $\beta = f/b$ are estimated using the \LCDM{} growth rate value, $f=\Omega_m^{0.55}$. The $\sigma_8$-values are known for the \LCDM{} and neutrino simulations and otherwise estimated as explained in the main text.}            
\label{table:sim_properties}      
\centering                         
\begin{tabular}{c c c c}        
\hline\hline                 
Simulation & $b$ & $\beta$ &$\sigma_8$\\    
\hline                        
   \texttt{lcdm}& $1.436$ & $0.3715$& $0.85$\\
   \texttt{fofr}& $1.349$& $0.3954$& $0.89$\\
   \texttt{015ev}& $1.484$ & $0.3595$ & $0.82$\\
   \texttt{06ev}& $1.739$ & $0.3067$& $0.73$\\
   \texttt{fofr\_015ev}& $1.397$ & $0.3818$ & $0.86$\\
   \texttt{fofr\_06ev}& $1.624$ & $0.3285$ & $0.76$\\
   \hline
   \texttt{lcdm\_small}& $1.271$ & $0.4146$ & $0.80$\\
   \texttt{fofr\_small}& $1.091$& $0.4830$ & $0.91$\\
\hline                                   
\end{tabular}
\end{table}

Looking at the bias values on linear scales, we can calculate an estimate of the linear halo bias. This is presented in Table~\ref{table:sim_properties}, together with $\beta$ and $\sigma_8$. The two latter are simulation parameters, $\beta$ being the reconstruction parameter of Sect.~\ref{sec:reconstruction} and $\sigma_8$ the amplitude of the linear matter power spectrum at a scale of $8\,h^{-1}\mathrm{Mpc}$. For the \LCDM{} and massive neutrino simulations, $\sigma_8$ is a known parameter obtained from the linear power spectrum calculated by \CLASS{}\footnote{\url{https://github.com/lesgourg/class_public}} as a part of the initial condition set-up (see Sect.~\ref{sec:ic}). For the simulations including modified gravity, an estimation of the linear $\sigma_8$-value was found by observing that $\sigma_{8, \CLASS}^{\nu}/\sigma_{8, CLASS}^{\Lambda\mathrm{CDM}} \approx \sigma_{8, \ANUBISIS}^{\nu}/\sigma_{8, \ANUBISIS}^{\Lambda\mathrm{CDM}}$ for the linear values calculated from \CLASS{} and the non-linear values calculated from \ANUBISIS. We then assumed that this relation also holds in the modified gravity context.

\subsection{Initial conditions}
\label{sec:ic}
The initial conditions used for the \ANUBISIS{} simulations were generated by following the same procedure as outlined in \cite{MNCCP}. In short, the linear matter power spectra and transfer functions were generated by \CLASS{} \citep{Lesgourgues2011, Blas2011}, and rescaled to $z=127$ through the \REPS{}\footnote{\url{https://github.com/matteozennaro/reps}} code \citep{Zennaro2017}, which takes the effects of massive neutrinos into account through a two-fluid description and by including radiation in the background evolution and implementing a scale-dependent growth rate. Positions and velocities for the CDM and neutrino particles were then generated by a version of the \ngenic{} code\footnote{\url{https://github.com/franciscovillaescusa/N-GenIC_growth}} that also has been modified to include the scale-dependence of the growth rate and growth factor in the presence of massive neutrinos.

For the initial conditions, we used $\Omega_b=0.049$ and $\Omega_c \simeq 0.27$. We kept the total matter density fixed at $\Omega_m\simeq 0.319$ by adjusting the ratio of $\Omega_c$ and the neutrino density parameter, $\Omega_{\nu} = \sum m_{\nu}/(93.14h^2\,\mathrm{eV})$. Besides this, we used the Hubble constant, $h=0.67$, the scalar spectral index, $n_s=0.9619$, the CMB temperature, $T_{\rm{CMB}}=2.7255\,\mathrm{K}$, and the amplitude of the primordial power spectrum, $A_s=2.215\times10^{-9}$, at the pivot scale $k_p=0.05\,\mathrm{Mpc}^{-1}$. This is the same cosmology as applied by \cite{MNCCP}.

We used the method outlined above to generate initial conditions for our \LCDM{} and massive neutrino simulations. For the simulations including \fofr{}-modified gravity, we used the outputs from the other simulations at $z=4$ as initial conditions. The deviations from GR are small at this stage, and only become more important at later times, as demonstrated for $z=3$ and $|f_{R0}|=10^{-5}$ by \cite{Zhao2011}. For the two independent \ISIS{} runs, the initial conditions were generated by \grafic{} \citep{Bertschinger2001}, with the parameters $\Omega_c=0.267$, $\Omega_b=0.045$, $h=0.719$, and $n_s=1.0$.

\section{Method}
\label{sec:method}
In this section, we detail the various codes and packages used to study our simulation data. We also provide an overview of the full analysis process.

\subsection{Halo finder}
\label{sec:halofinder}
To identify halos in our simulations, we used the \rockstar{}\footnote{\url{https://bitbucket.org/gfcstanford/rockstar/src/main/}} halo finder \citep{Behroozi2012}. \rockstar{} locates halos in phase space by applying the 3D friends-of-friends (FOF) method to pinpoint overdense regions. For each of the groups created in the FOF procedure, the linking length (the characteristic length scale grouping the particles together) is reduced progressively, so that subgroups emerge in a hierarchical structure. Seeds are then placed in the lowest substructures and particles are assigned to the halo seed within the shortest phase-space distance. From this, the relationship between host and subhalos is computed and unbound particles are removed. Finally, halo properties are calculated.  

When applying \rockstar{} to our simulations, we read the gadget files directly to obtain the CDM particle properties and the cosmological parameters. The force resolution was set to be approximately the smallest distance resolved by the simulation, and the linking length was set to $l = 0.28\,h^{-1}\mathrm{Mpc}$, which is $0.2$ of the mean inter-particle separation. In the end, we removed the subhalos and were left with a host halo catalogue. As shown in \cite{MNCCP}, \ANUBIS{} underestimates the number of halos, especially at low mass. This can be improved upon by increasing the particle density but was not done for the simulations presented in this paper. Because of this, we made no further mass cuts to the halo catalogue. 

\subsection{Void finder}
\label{sec:voidfinder}
When identifying voids within the halo catalogue of a simulation, we must first define a void numerically. In this paper, we apply two different void definitions, spherical voids and voxel voids. 

For the spherical void detection, we used the void finder module implemented in \pylians{}\footnote{\url{https://github.com/franciscovillaescusa/Pylians3}}, which is based on the method detailed in \cite{Banerjee2016}. Here, the algorithm is provided with a range of radii corresponding to spherical regions of various sizes to be investigated, along with an overdensity threshold given by $\Delta = \rho/\bar{\rho} -1$. Starting with the largest radius, the simulation box is divided into a grid and the density profile is calculated in each voxel and smoothed on a scale corresponding to the given radius. Voxels with densities below the threshold are identified and sorted. Starting with the lowest density voxel, all the voxels around it within the given radius are considered and added as a part of the void. This is then repeated for the next to lowest density voxel and so on. If any of the voxels within the corresponding radius are already assigned to another void, the new void is rejected. Once this is done for all the voxels below the density threshold, the algorithm moves on to the next to largest radius provided, smooths the density field at that scale, and once again proceeds as explained above. This is repeated for all the radii given initially, and the resulting voids identified are spherical regions of these specific sizes, with the centre at the lowest density voxel. In our case, we set $\Delta = -0.75$ and provided the algorithm with 47 void radii between $16-82\,h^{-1}\mathrm{Mpc}$.

For detecting voids by using the voxel void definition, we employed the \revolver{}\footnote{\url{https://github.com/seshnadathur/Revolver}} void finder \citep{Nadathur2019revolver}. Here, the voxel voids are defined by using a watershed method. The simulation box is divided into a grid and local minima of the density field are located. Around each of these minima, surrounding voxels with increasing overdensity are added to the pool up until a voxel with a lower overdensity than the one previously added is discovered. Using this definition, the identified voids may have any shape, as opposed to the spherical void definition. The centre of the void again lies within the voxel with the lowest density. The \revolver{} void finder can also perform reconstruction when provided with a halo catalogue given in redshift space. This is based on linear theory, as described in Sect.~\ref{sec:reconstruction}, and requires the halo bias, $b$, and growth rate, $f$, as input parameters. For a more detailed description of the spherical and voxel void definitions, along with other void-identifying algorithms, see \cite{Massara2022}.

\subsection{Covariance matrix}
\label{sec:covmat}
Each of our simulation boxes only has one realisation. To obtain an estimate of the uncertainty in the redshift space void-halo CCF calculated from the simulation data, we attained the covariance matrix through a Jackknife method. We applied the Jackknife estimator as implemented in \pycorr{}\footnote{\url{https://github.com/cosmodesi/pycorr}}, following \cite{Mohammad2022}.

The \ANUBISIS{} simulation boxes were equally divided into $n=512$ sub-boxes\footnote{We ran tests using both less ($216$) and more ($2744$) jackknife samples and found similar estimates for the covariance matrix.}. Individual Jackknife realisations were then made by calculating the correlation function for the full volume with one of the sub-boxes removed at a time. This lead to $n$ Jackknife realisations, each with a volume fraction $(n-1)/n$ of the original volume. Based on this, each element in the covariance matrix, $\tens{C}_{ij}$, is given by
\begin{align}
    \tens{C}_{ij} = \frac{n-1}{n}\sum_{k=1}^n\big[\xi_i^k - \bar{\xi}_i\big]\big[\xi_j^k - \bar{\xi}_j\big],
\end{align}
where 
\begin{align}
    \bar{\xi}_i = \frac{1}{n}\sum_{k=1}^n\xi_i^k
\end{align}
is the mean estimate from all the $n$ Jackknife realisations.

It is important to note that since we only have one simulation of each individual case, and we used the data from the simulations to calculate $\xi^r$, $v_r(r)$, and $\sigma_{v_\parallel}$, the model and data $\xi^s$ are correlated. Ideally, the input parameters should be calculated from the average of many mock simulations, as is done in for example \cite{Woodfinden2022} and \cite{Nadathur2019}, to avoid this problem, along with the issue of the uncertainty associated with only having one measurement of $\xi^r$. One way to deal with the correlated errors is to calculate the covariance matrix for the CCF of the difference between the model and the data, $\xi^s_{\mathrm{diff}} = \xi^s_{\mathrm{model}} - \xi^s_{\mathrm{data}}$. A more detailed explanation and demonstration of the effects of this approach can be found in Appendix A of \cite{Radinovic2023}. We did not include this step in our analysis as we focus more on the comparison between the simulations as opposed to reducing the statistical error for a single simulation.

\subsection{Analysis pipeline}
\label{sec:pipeline}
For the \ANUBISIS{} simulation data, the CDM and neutrino particles were given separate identifiers. It is the CDM particle information that goes into the procedure detailed below unless otherwise stated. 

First, the particle data were given to the \rockstar{} halo finder, which identified halos in the simulation boxes as detailed in Sect.~\ref{sec:halofinder}. This provided both the positions and velocities of the halos, enabling us to put the halos into redshift space if needed. The halo catalogues were then given to the \revolver{} void finder, which identified voxel voids in between the halos, as explained in Sect.~\ref{sec:voidfinder}. In order to test how well reconstruction (Sect.~\ref{sec:reconstruction}) performs in the case of massive neutrinos and modified gravity, this step was not only performed directly on the real space halo catalogue but also with the redshift space halo catalogue as input. We performed the reconstruction step by applying \revolver{}, which solves Eq.~(\ref{eq:recon}) through an iterative fast Fourier transform algorithm \citep{Burden2014,Burden2015}. This is the same algorithm used for BAO reconstruction (e.g. \cite{Gil-marin2020,Bautista2020}). We then had two void catalogues, one directly identified in real space, and one in reconstructed real space, both found by the voxel void definition. We then performed the same void finding with the spherical void definition, as detailed in Sect.~\ref{sec:voidfinder}, both with the real space and reconstructed real space halo positions. After obtaining void catalogues in real and reconstructed real space, for the voxel and spherical void definitions, we performed a radius cut at $21\,h^{-1}\mathrm{Mpc}$ for all catalogues, to eliminate void discoveries that coincide with the average spacing between halos. For each catalogue, the remaining voids with their coinciding halos were then stacked on top of each other to obtain an `average' void with more statistically robust properties and spherical symmetry. From the halo distribution in and around these voids, we calculated the density profile, mean velocity profile, and LOS velocity dispersion. The latter was computed as detailed by \cite{Radinovic2023}. These are all ingredients needed to calculate the void-halo CCF in redshift space as shown in Eq.~(\ref{eq:streaming_coord_change}). 

We want to compare the modelled void-halo CCF in redshift space with the void-halo CCF calculated directly from the simulation data. In order to obtain the latter we used the correlation function estimation wrapper, \pycorr{}, which currently utilises the \corrfunc{}\footnote{\url{https://github.com/manodeep/Corrfunc}} CCF engine \citep{Sinha2019,Sinha2020}. Provided the redshift space halo catalogue, real space void catalogue, and randomly distributed reference catalogues, \pycorr{} calculates the void-halo CCF of the simulation data as we would observe it. This is done using the Landy-Szalay estimator \citep{Landy1993},
\begin{align}
    \xi(r,\mu) = \frac{D_1D_2(r,\mu)-D_1R_2(r,\mu) - D_2R_1(r,\mu) + R_1R_2(r,\mu)}{R_1R_2(r,\mu)},
\end{align}
where $D$ is the simulated data catalogues, $R$ is the random catalogues, and 1 and 2 are the halos and voids. We used 50 bins between $0-150\,h^{-1}\mathrm{Mpc}$ for $r$, the spatial distance between the pairs, and 200 bins from $-1$ to $1$ for $\mu$, the angular separation. For the random catalogues, we used 50 times as many halos and voids as the simulations. \pycorr{} also has an inbuilt jackknife estimator, which allowed us to calculate the covariance matrix of the CCF, as described in Sect.~\ref{sec:covmat}. We calculated the void-halo CCF using both the voids identified in real space and in reconstructed real space. 

Having the simulated CCF, we used \victor{}\footnote{\url{https://github.com/seshnadathur/victor}} to calculate the theoretical model and compare it to the data, both in the real and reconstructed case. To do so, \victor{} requires $\xi^s$ from the simulations, a covariance matrix, and $\xi^r$, $v_r(r)$, $\Delta_\mathrm{DM}$, and $\sigma_{v_\parallel}$ as input to the model. \victor{} also assesses the goodness of fit between the data and the model, and provides a $\chi^2$ value upon request. 

Through an interface with \cobaya{} (code for bayesian analysis)\footnote{\url{https://github.com/CobayaSampler/cobaya}} \citep{Torrado2019,Torrado2021}, we can also use \victor{} to perform MCMC fits of the parameters in the void-halo CCF model. We assumed a Gaussian form of the likelihood 
\begin{align}
    \log\mathcal{L} = -\frac{1}{2}\Big(\mathbf{\xi}^s_{\mathrm{model}} - \mathbf{\xi}^s_{\mathrm{data}}\Big)\tens{C}^{-1}\Big(\mathbf{\xi}^s_{\mathrm{model}} - \mathbf{\xi}^s_{\mathrm{data}}\Big)^\mathrm{T},
\end{align}
along with flat priors for the parameters $f\sigma_8$, $\sigma_{v_\parallel}$, $\beta$, and $\epsilon$, encompassing the fiducial values. In the case where we used the actual real space data from the simulations, $\beta$ was not included. In the case of reconstruction, $\beta$ was allowed to vary. When comparing CCF data to theory, reconstruction was only performed for the fiducial $\beta$-value of each simulation. However, to allow for $\beta$ to vary for the MCMC fits, we performed reconstruction, void finding, and the CCF calculation repeatedly for 11 different $\beta$-values, which were provided to \victor{} to perform the fit. Ideally, covariance matrices should also be computed in each reconstruction case, as it depends on $\beta$. This is a time-consuming task, and as a first approach, we instead assumed a fixed $\beta = \beta^{\mathrm{fiducial}}$ for all the covariance matrices. We also kept the input $\xi^r$ equal to the one calculated for $\beta^{\mathrm{fiducial}}$ in all cases, meaning that the $\beta$-dependence only shows up in $\xi^s$. This is the same approach taken by \cite{Radinovic2023}.

\section{Results}
\label{sec:results}
In this section, we present the results of our analysis. In addition, we discuss the implications of our results.

\subsection{Void abundance}
\begin{figure}
    \hspace{-0.3cm}
    \includegraphics{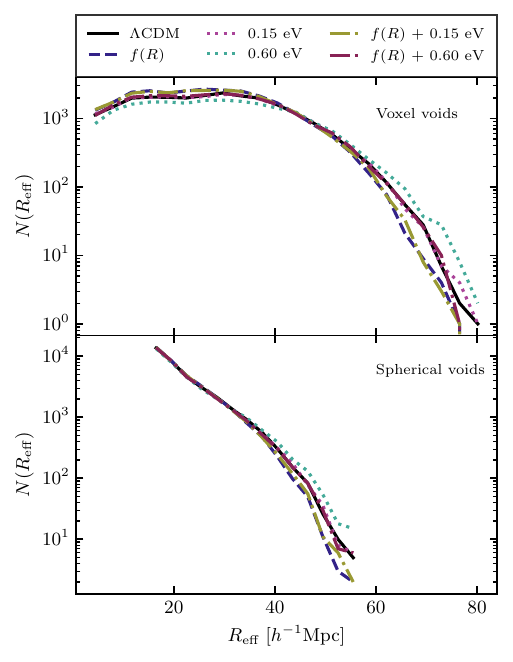}
    \caption{Histogram of effective void size for the various \ANUBISIS{} simulations. The voxel voids display a notable difference in the number of voids in the $\sim10-40\,h^{-1}\mathrm{Mpc}$ range. Both void definitions also show a slight opposite abundance difference amongst the largest voids in the simulation catalogues.}
    \label{fig:nr_own_sims_hist}
\end{figure}

First, we take a look at the general void population identified in all the \ANUBISIS{} simulations. In Fig.~\ref{fig:nr_own_sims_hist} we show the abundance of voids depending on the effective void radius. At the top, the voxel void definition has been used to find the voids, and at the bottom, the spherical void definition. For the voxel voids, the difference in the number of voids is clearly apparent in the approximate range $10-40\,h^{-1}\mathrm{Mpc}$. For the spherical voids, however, this is not the case. If we zoom in, the same ordering is present, but this is not apparent at first glance due to the high number of voids identified in the given range. The spherical void finding algorithm, as detailed in Sect.~\ref{sec:voidfinder}, identifies voids by smoothing the field for a given top-hat radius and declaring spheres with density below a certain threshold as voids. For small radii, this results in a large amount of voids, some of which might be the result of shot noise, for all simulations. 

    \begin{figure}
   \hspace{-0.4cm}
            {\includegraphics[scale=1.05]{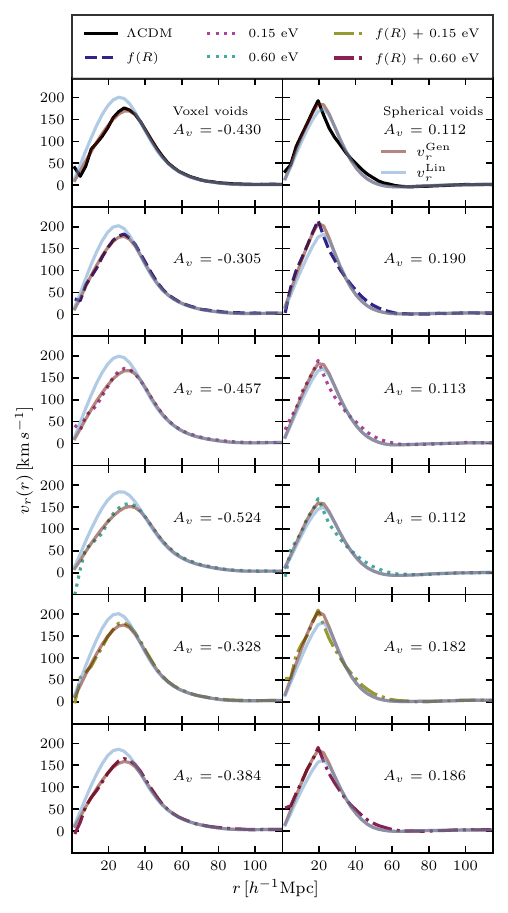}}
      \caption{Radial void velocity profiles for all \ANUBISIS{} simulations together with the linear velocity profile and a fit to the more general velocity profile, presented in Sect.~\ref{sec:vel_modelling}. The voxel and spherical void definitions are shown in the left and right columns respectively. The $A_v$-parameter in each panel shows the best-fit value of the general profile. 
      }
         \label{fig:vr_sim_theory}
   \end{figure}

For the \fofr{} and \fofr{} dominated \fofr{} $+\, 0.15\,\mathrm{eV}$ simulations, we see an increase in the number of voids within the $10-40\,h^{-1}\mathrm{Mpc}$ effective radius range, compared to \LCDM{}. This is due to the fifth force enhancing gravity in these regions, resulting in a more effective `evacuation' of the areas. \cite{Li2012} also reports a higher number of the large voids in their \fofr{} simulations, although it should be noted that their maximum void size is around $15\,h^{-1}\mathrm{Mpc}$ due to their smaller, but higher resolution, simulation boxes. 

The most massive neutrino simulation, $0.60\,\mathrm{eV}$, shows the opposite behaviour. Here, the amount of voids in the given size range is suppressed, compared to \LCDM{}. This is a result of the neutrinos slowing down the clustering and thereby the evolution of the voids towards lower densities. \cite{Massara2015} also reports fewer large voids in their massive neutrino simulations, compared to \LCDM{}, where large in their case is in the range $20-40\,h^{-1}\mathrm{Mpc}$. For very large voids, $R_\mathrm{eff}\gtrsim45\,h^{-1}\mathrm{Mpc}$, we see a turn-around of the ordering for both void definitions. \cite{Cai2015} also found a higher amount of very large radii ($\gtrsim 25\,h^{-1}\mathrm{Mpc}$) voids in their regular GR simulations, compared to \fofr{}. Based on investigation they suggest that this is a result of the largest voids being less empty in the \fofr{} case, as the enhanced gravity inside makes it easier for halos to form, compared to \LCDM. When looking at a big region in \LCDM{} versus a big region in the \fofr{} simulation, the number density of halos would be lower for \LCDM{}, making it easier to pass the void identification criteria. The opposite could then be argued for the massive neutrino simulations. It should be noted that the differences observed for very large voids in Fig.~\ref{fig:nr_own_sims_hist} are enhanced due to the logarithmic scale, and could also be affected by the small sample size.

For the \LCDM{} simulation, we find in total $N_{\mathrm{halo}} \approx 1.3\times10^6$. From this, we chose a radius cut for the voids as $R_{\mathrm{eff}}^{\mathrm{cut}} = 1.5(N_{\mathrm{halo}}/V_{\mathrm{box}})^{-1/3} \approx 21\,h^{-1}\mathrm{Mpc}$. This was applied to the void catalogues to make sure that we exclude what are simply empty regions in between halos in the simulations, and not actual voids.

\subsection{Velocity profile}
\label{sec:result_vel_profile}

   \begin{figure}
    \hspace{-0.4cm}
            {\includegraphics[scale=1.05]{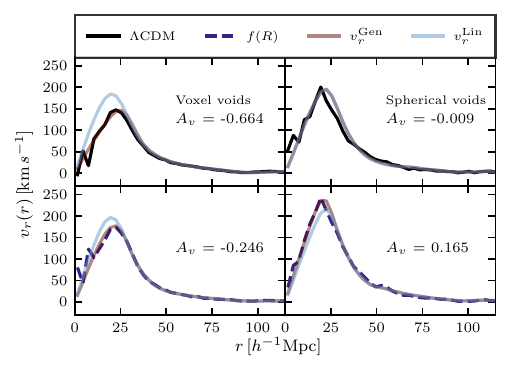}}
      \caption{Radial void velocity profiles for the \texttt{fofr\_small} and \texttt{lcdm\_small} simulations together with the linear velocity profile and a fit to the more general velocity profile. The voxel and spherical void definitions are shown in the left and right columns respectively. The $A_v$-parameter shows the best-fit value for the general profile.}
         \label{fig:f4_vr_sim_theory}
   \end{figure}

When using galaxy surveys to identify voids, the velocity profile needed for the CCF model is typically modelled by the linear velocity profile or estimated from simulations. Because of this, we want to study how the simulated velocity profiles for various cosmologies compare to the linear model. In Fig.~\ref{fig:vr_sim_theory}, the mean radial outflow void velocity profile for each \ANUBISIS{} simulation is shown for both the voxel and spherical void definitions. The individual profiles are compared to the theoretical linear velocity profile, as shown in Eq.~(\ref{eq:vr_lin}), and the more general velocity profile, as shown in Eq.~(\ref{eq:general_vr}). For the linear velocity profile, the growth rate, $f=\Omega_m^{0.55}$, corresponding to the expected \LCDM{} value, was applied for all simulations. In doing so, we ignore the scale dependence of the growth rate in the \fofr{} and massive neutrino cosmologies and investigate whether or not this leads to biased results. For the general profile, the best-fit values of the $A_v$-parameter are given in each panel of Fig.~\ref{fig:vr_sim_theory}. These were obtained through the least squares method. 

The general velocity profile is, with the best fit $A_v$-parameter values, by construction, a good match to the velocity profiles found in the simulation data. For the voxel voids, the best-fit parameter value is consistently higher for \fofr{} and decreases with increasing neutrino mass, compared to \LCDM{}. This is expected as a result of higher velocities in the \fofr{} case, and lower in the massive neutrino case. The mixed simulations lie somewhere in between. For the spherical voids, we do not see the same pattern for the neutrino masses and the value of $A_v$. However, for the spherical voids, the general velocity profile does not fit as well as for the voxel voids at higher radii. When estimating $A_v$, this must be taken into account, in addition to the peak of the velocity profile. If we could observe the velocity profile directly over a large void sample, the best-fit $A_v$-value could be an interesting parameter to explore as a cosmological probe. Observing the velocity profile directly is not possible in traditional galaxy surveys, which is why it is currently modelled or estimated from simulations.

   \begin{figure}
    \hspace{-0.2cm}
    \includegraphics{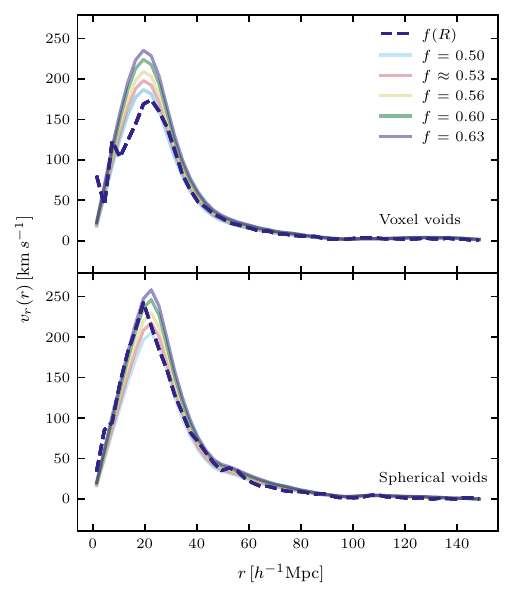}
    \caption{Linear velocity profile, compared to the \texttt{fofr\_small} simulation, for various values of the growth rate, $f$. The regular linestyle and colour show the velocity profile for the simulation. For \LCDM{} we have $f = \Omega_M^{0.55}\approx 0.53$. 
    }
    \label{fig:f4_v_vary_f}
\end{figure}

From visual inspection, it is clear that the theoretical linear velocity profile is not a good fit for the data close to the void centre. This is due to a sparse tracer sample in this region and also depends on the void centre definition \citep{Massara2022}. It affects the voxel void definition more than the spherical one. In the voxel void case, the linear model overestimates the clustering. We can see that this is somewhat compensated for in the \fofr{} simulations, due to higher velocities. This is even more visible for the pure \ISIS{} simulations, as shown in the lower left panel of Fig.~\ref{fig:f4_vr_sim_theory}. For the spherical voids, this is less of an issue, although the linear model slightly underestimates the clustering. In the latter case, this could be compensated by adopting a different value of the growth rate, $f$, in the linear theory model, as demonstrated in Fig.~\ref{fig:f4_v_vary_f}. For the voxel voids, increasing the value of $f$ in line with expectations (e.g. \cite{Mirzatuny2019}) only increases the discrepancies.

Altogether, it is evident that the linear velocity profile is not a drastically worse fit for the \fofr{} or massive neutrino simulations. Before we can address any differences due to cosmology, however, we must make sure that a sparse tracer sample does not affect our model-data comparison. As it stands now, applying the linear velocity profile to the CCF modelling results in modifications similar to changes in $f\sigma_8$, due to the offset induced by the number of tracers \citep{Massara2022}. In addition, we should point out that even if the velocity modelling was improved to forgo this issue, the modifications expected from the $|f_{R0}|=10^{-5}$ \fofr{} simulation and the most massive neutrino simulation are mostly on scales smaller than the average void size. 

\begin{figure}
    \centering
    \includegraphics{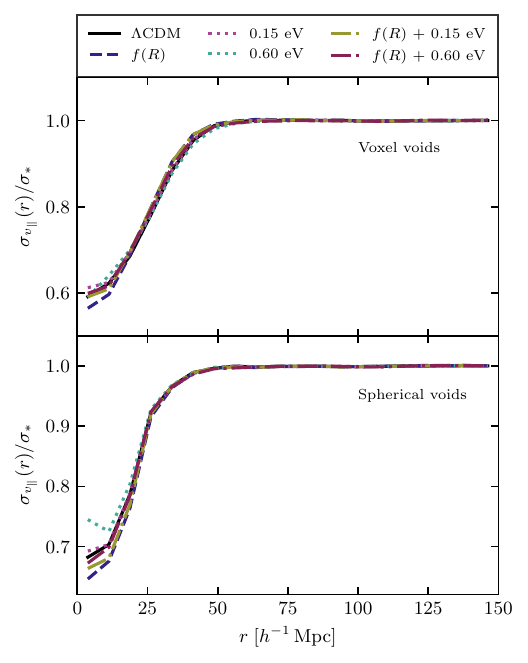}
    \caption{LOS velocity dispersion scaled by $\sigma_*$ as given in Table~\ref{table:sigma_star}. The voxel and spherical void definitions are shown in the upper and lower panels respectively. All the simulated cosmologies result in similar profile shapes. The largest differences are located towards the void centres, which are dominated by shot noise.}
    \label{fig:v_disp_own_sims}
\end{figure}

      \begin{figure*}
   \resizebox{\hsize}{!}
            {\hspace{-0.2cm}\includegraphics{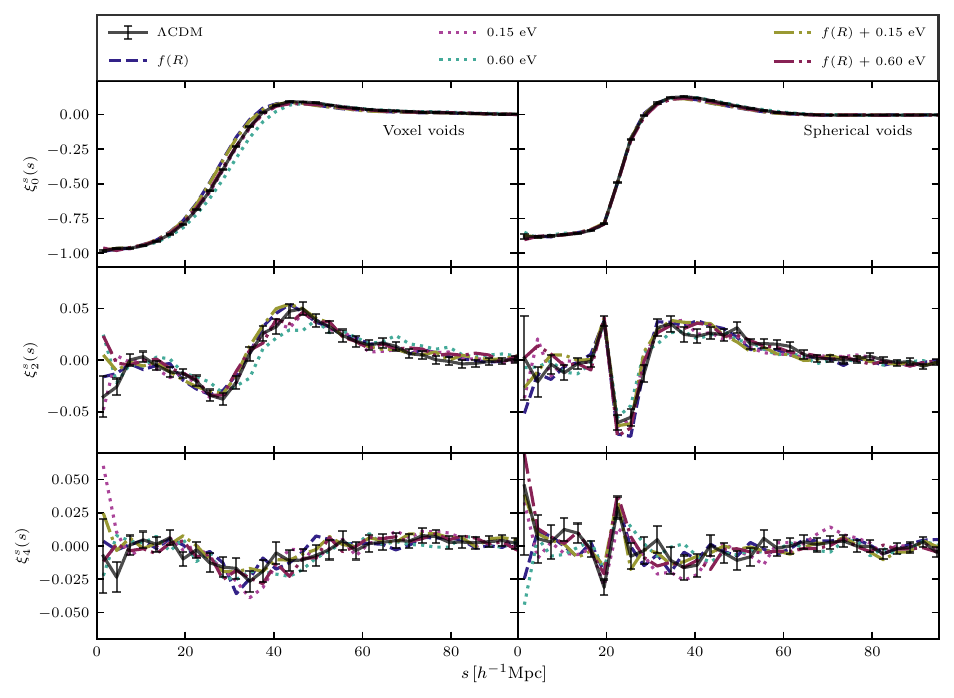}}
      \caption{Void-halo CCF for the \ANUBISIS{} simulation data. The voxel and spherical void definitions are shown in the left and right columns respectively. The error bars are estimated using a Jackknife technique and are only shown for the \LCDM{} case for a tidier visual representation of the data. 
      }
         \label{fig:ccf_jackknife}
   \end{figure*}
\subsection{Velocity dispersion}

\begin{table}
\caption{Average value of the LOS velocity dispersion, $\sigma_*$, for $r\geq75\,h^{-1}\mathrm{Mpc}$ in the case of voxel voids (VV) and spherical voids (SV).}             
\label{table:sigma_star}      
\centering                          
\begin{tabular}{c c c}        
\hline\hline                 
Simulation &$\sim\sigma_*^{\mathrm{VV}}\,[\mathrm{km}\,\mathrm{s}^{-1}]$ & $\sim\sigma_*^{\mathrm{SV}}\,[\mathrm{km}\,\mathrm{s}^{-1}]$ \\    
\hline                        
   \texttt{lcdm}& $334.4$ & $334.3$ \\
   \texttt{fofr}& $356.7$ & $356.9$\\
   \texttt{015ev}& $329.2$ & $328.8$ \\
   \texttt{06ev}& $300.8$ & $300.6$\\
   \texttt{fofr\_015ev}& $351.0$ & $351.0$ \\
   \texttt{fofr\_06ev}& $320.8$ & $320.3$\\
   \hline
   \texttt{lcdm\_small}& $319.4$ & $319.6$ \\
   \texttt{fofr\_small}& $388.1$ & $387.9$\\
\hline                                   
\end{tabular}
\end{table}

The CCF theory presented in Sect.~\ref{sec:ccf_theory} depends on the velocity dispersion of the void velocity profile. For all the \ANUBISIS{} simulations, Fig.~\ref{fig:v_disp_own_sims} shows this quantity for both the voxel and spherical void definitions, scaled by the average amplitude beyond $75\,h^{-1}\mathrm{Mpc}$, $\sigma_*$. The velocity dispersion profile has the same shape for all the various cosmologies, but a different amplitude, $\sigma_*$, as presented in Table~\ref{table:sigma_star}. The relative values of the amplitudes coincide with expectations. The halos in and around the void in the \fofr{} simulation feel a stronger gravitational pull towards the edge, compared to the \LCDM{} case. In general, there is also more clustering which leads to higher velocities. The relative values of the amplitudes in the \fofr{} and \LCDM{} cases coincide well with the findings of \cite{Fiorini2022}. The opposite behaviour is seen for the massive neutrino simulations, where less clustering results in lower velocities. However, although these differences can be calculated from the simulated data, they can not readily be observed.

\subsection{Void-halo CCF: data versus model}
We are interested in the void-halo CCF of the various simulated cosmologies and how well the theory presented in Sect.~\ref{sec:ccf_theory} reproduces the simulation data. Figure~\ref{fig:ccf_jackknife} shows the monopole, quadrupole, and hexadecapole of the void-halo CCF for the \ANUBISIS{} simulations. The left and right columns display the results for the voxel and spherical void definitions respectively. The voids were identified from the real space halo catalogue, and the redshift space halo positions were calculated directly from the real space positions and corresponding LOS velocities.

      \begin{figure}
           \hspace{-0.4cm} {\includegraphics[scale=1.05]{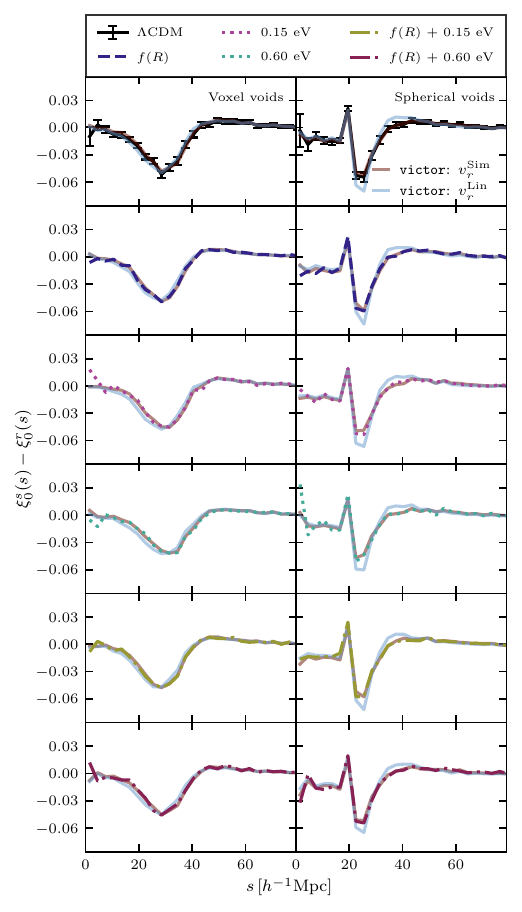}}
      \caption{Difference between the redshift and real space monopole of the void-halo CCF from the \ANUBISIS{} simulations vs. theory calculated by \victor. The results are obtained by using the voxel and spherical void definitions in the left and right columns respectively. Two theory results from \victor{} are included, one where the velocity profile is given by linear theory and one where it is given by a template. The template used in this case is the velocity profile calculated from the simulation data.}
         \label{fig:ccf_theory_jackknife_xi0}
   \end{figure}
   
For the monopole, the differences between the simulations are clearer for the voxel void definition, as was also the case earlier when looking at the void abundance. The shape of the monopole is closely related to the shape of the density profile around the void, as both in essence describe how halos are distributed around the void centre as a function of radius. The \fofr{} and \fofr{} $+\,0.15\,\mathrm{eV}$ simulations show the narrowest profiles (most evolved) and the $0.6\,\mathrm{eV}$ simulation the broadest (least evolved), while the others fall somewhere in between. From the void abundance histogram in Fig.~\ref{fig:nr_own_sims_hist}, we saw that the \fofr{} case has more voids in the $10-40\,h^{-1}\mathrm{Mpc}$ effective radius range and the $0.6\,\mathrm{eV}$ has less, compared to \LCDM{}. This further explains what we see here in terms of the shapes of the profiles, as the monopole is based on the overall averaged void from each simulation box. For the voxel void definition, the central value goes towards $-1.0$, while for the spherical void definition, we see a very flat core stabilising at around $-0.88$. This is again a result of the void-finding algorithm, where a sphere of a given radius with a smoothed density below $-0.75$ is classified as a void. As mentioned before, this leads to the detection of a large amount of small voids, and in addition, most of the detected voids have central densities that are larger than $-1.0$.  

The quadrupole is most easily studied for the voxel void definition. Here, we see, in all cases, the expected smooth shape resulting from RSDs as a consequence of halos inside the void moving towards the small overdensity at the edge, but also halos outside the void moving back towards the void edge. For the spherical voids, we instead see a sharp feature proceeding the edge of the void. The difference between the quadrupoles of the two void definitions is due to the shapes of the void density profiles and the presence of a velocity dispersion. This is further explained through the use of a toy model in Appendix~\ref{sec:quad_shape}. Within the error bars, the quadrupole for each simulation is not easily distinguishable. The most massive neutrino simulation shows signs of lower peaks due to overall lower velocities for the voxel void definition, but higher resolution simulations should be performed in order to confirm this. 

       \begin{figure}
           \hspace{-0.2cm}\includegraphics[scale=1.05]{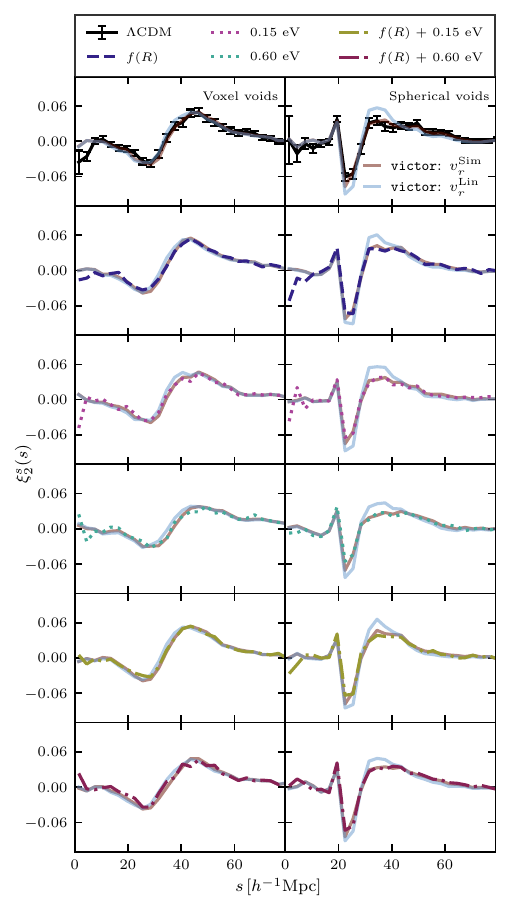}
      \caption{Quadrupole of the void-halo CCF from the simulations vs. theory calculated by \victor. Layout and input as explained for Fig.~\ref{fig:ccf_theory_jackknife_xi0}.}
         \label{fig:ccf_theory_jackknife_xi2}
   \end{figure}

The hexadecapole is, from the theoretical model, expected to be small \citep{Cai2016}. This is the case for all simulations and both void definitions, except for a little dip in the voxel void $\xi_4^s$ around $s\sim 35\,h^{-1}\mathrm{Mpc}$. This small signal in the hexadecapole shows why it can still be important to include this multipole when fitting the model to the data to obtain cosmological parameters, instead of assuming it to be zero.

    \begin{figure}
            \hspace{-0.2cm}{\includegraphics[scale=1.05]{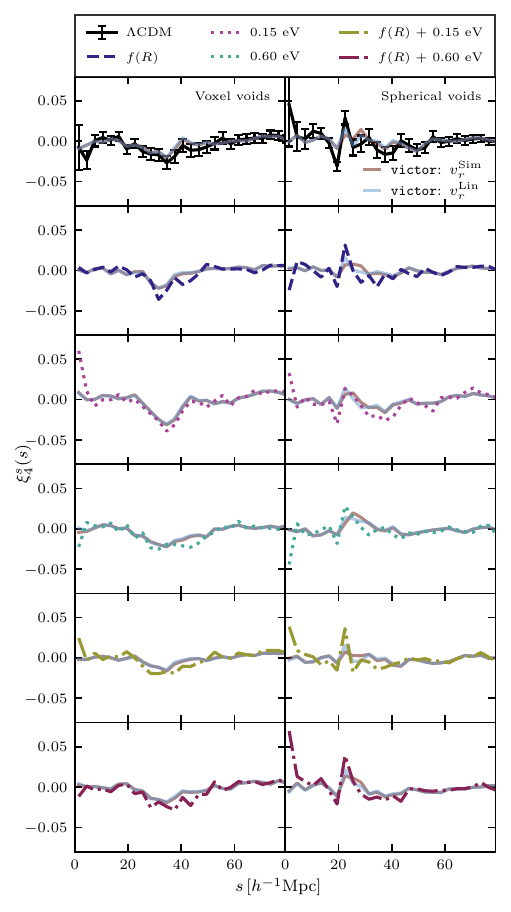}}
      \caption{Hexadecapole of the halo-void CCF from the simulations vs. theory calculated by \victor. Layout and input as explained for Fig.~\ref{fig:ccf_theory_jackknife_xi0}.} 
         \label{fig:ccf_theory_jackknife_xi4}
   \end{figure}

Figures~\ref{fig:ccf_theory_jackknife_xi0}, \ref{fig:ccf_theory_jackknife_xi2}, and~\ref{fig:ccf_theory_jackknife_xi4} respectively display the monopole, quadrupole, and hexadecapole of the void-halo CCF in redshift space as calculated directly from the \ANUBISIS{} simulation data, compared to its theoretical value calculated by \victor{} as outlined in Sect.~\ref{sec:ccf_theory}. The simulation data follow the regular colour and linestyle pattern, while the theory is presented with full brown and blue lines. The brown line is the relevant theoretical multipole as calculated by \victor{}, with the velocity profile from the simulation data given to the model as a velocity template, and the blue line is the same, only with the linear velocity profile as displayed in Eq.~(\ref{eq:vr_lin}) given as input. For each simulation, void definition, and multipole, only the \LCDM{} simulation data line has error bars. This is to illustrate their magnitude but otherwise reduce cluttering. 

The monopole data-theory comparison in Fig.~\ref{fig:ccf_theory_jackknife_xi0} is presented through the difference between the redshift space and real space monopole. This is done to better display the discrepancies between the data and the two models. For all simulations and both void definitions, we see good agreement. There is a preference towards inputting the velocity profile from the simulations in the model, compared to using the linear velocity profile. This is expected simply because we are providing the model with the actual data that we want to reproduce. We already saw, in Sect.~\ref{sec:result_vel_profile}, that the linear velocity profile does not reproduce the velocity profile directly calculated from the halos in the simulations very well. However, if this study was performed on observational data, we would not have had access to the exact mean outflow velocity profile. We could then either use the linear approximation or make estimates through simulations. 

       \begin{figure}
       \hspace{-0.2cm}
        \includegraphics[scale=1.05]{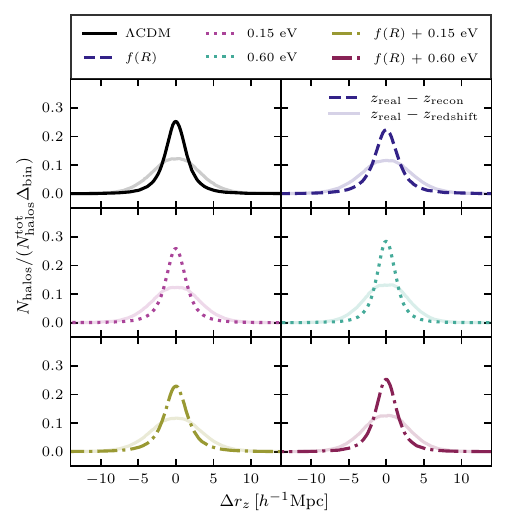}
      \caption{Difference in the LOS coordinate between real space and reconstructed real space for the \ANUBISIS{} simulations shown in the regular linestyle and colour. The weaker full line of the same colour shows the difference between the LOS coordinate in real and redshift space.} 
         \label{fig:recon_hist}
   \end{figure}

The quadrupole data-theory comparison in Fig.~\ref{fig:ccf_theory_jackknife_xi2} is the most interesting to study, as it is the RSDs observed here that, through modelling, can give us an estimate of the $f\sigma_8$-value. It is again evident from visual inspection that the model reproduces the data well for all simulations. For the voxel void definition, applying the linear velocity profile as an approximation gives results within the error bars for all simulations. For the spherical void definition, the consequence of applying the linear velocity profile is more significant, even though the difference between the data and linear theory velocity profiles in Fig.~\ref{fig:vr_sim_theory} is larger for the voxel void definition. This could be partly because even though the linear and simulation velocity profiles deviate more towards the void centre for the voxel voids, compared to the spherical voids, they actually coincide better for $r\gtrsim30\,h^{-1}\mathrm{Mpc}$. This is also where we see some of the larger discrepancies between the spherical void simulation quadrupole and the modelled quadrupole with the linear velocity as input. Even though we see a difference in the model fit for the two void definitions, there is no significant difference in the performance of the model between the various simulated cosmologies. Ideally, we need higher resolution simulations along with more accurate velocity modelling in order to quantify this.

Figure~\ref{fig:ccf_theory_jackknife_xi4} shows that also the hexadecapole is well described by the model for both void definitions. This further encourages including it in analysis when it is available.

\subsection{Reconstructed void-halo CCF: data versus model}

The reconstruction process, detailed in Sect.~\ref{sec:reconstruction}, assumes linearity and a constant growth rate which might affect its performance when modified gravity or massive neutrinos are introduced. Figure~\ref{fig:recon_hist} shows how reconstruction performs for each individual \ANUBISIS{} simulation. The lines, following the regular colour and style pattern, show a histogram of the difference between the LOS coordinate (defined as the $z$-direction in our simulation boxes) in real space and reconstructed real space. The fainter line, in the same colour, shows a histogram of the difference in the LOS coordinate in real space and redshift space, before any reconstruction is performed. The reconstruction process illustrated in this figure is executed with the fiducial $\beta$-values presented in Table~\ref{table:sim_properties}.

   \begin{figure}
    \centering
    \includegraphics{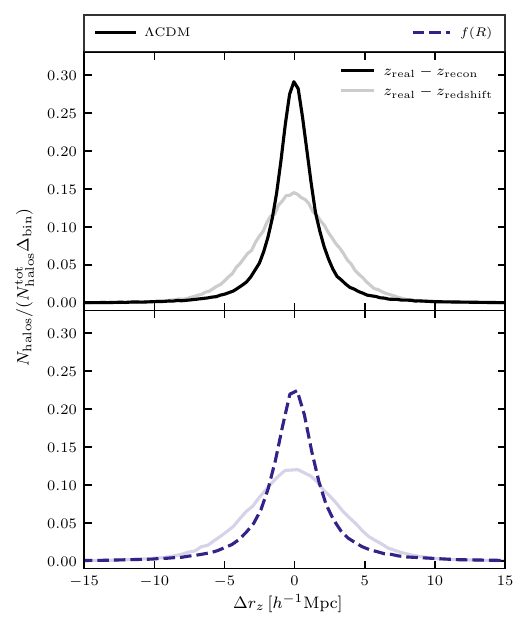}
    \caption{Difference in the LOS coordinate between real and reconstructed real space and real and redshift space for the \texttt{fofr\_small} and \texttt{lcdm\_small} simulations.}
    \label{fig:f4_recon_hist}
\end{figure}

       \begin{figure}
            \hspace{-0.2cm}{\includegraphics[scale=1.05]{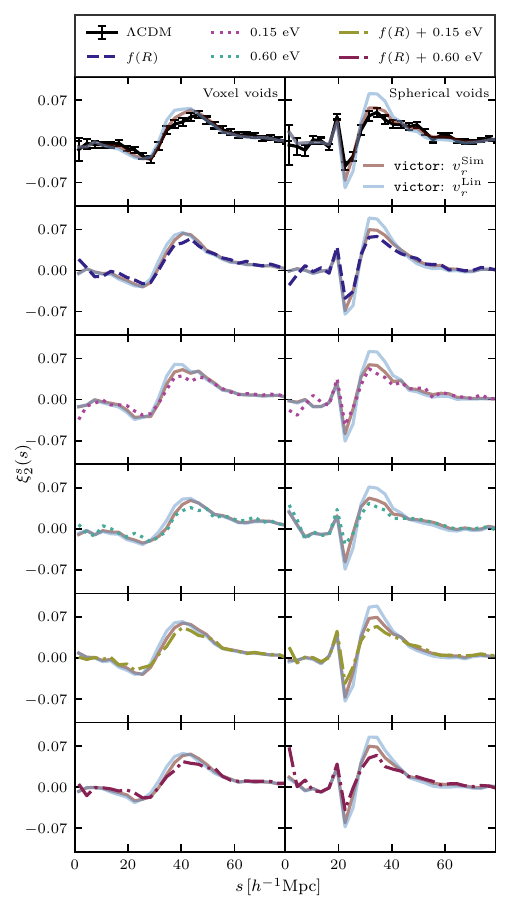}}
      \caption{Quadrupole of the reconstructed void-halo CCF from the simulations vs. theory calculated by \victor. The real space CCF used as theory input is calculated from the reconstructed real space halo catalogue and the voids identified in it. The layout follows Fig.~\ref{fig:ccf_theory_jackknife_xi0}.
      }
         \label{fig:ccf_theory_jackknife_xi2_recon}
   \end{figure}

For each simulation, the histogram of the difference between real and redshift space is always broader than that for the difference between real and reconstructed real space. This tells us that performing reconstruction has indeed led us closer to the actual real space values. The method is tested on \LCDM{} simulations by, for example, \cite{Nadathur2019recon,Woodfinden2022} and \cite{Radinovic2023}, and we achieve similar results in our case. The non-\LCDM{} simulations again obtain comparable results to the \LCDM{} reference. Although the \fofr{} dominated ones (\fofr{} and \fofr{} $+\,0.15\,\mathrm{eV}$) seem to perform slightly worse, and the heaviest neutrino mass, $0.6\,\mathrm{eV}$, slightly better. The reconstruction method is based on linear theory, and one possible explanation for this could therefore be that the massive neutrino simulations, with their suppressed clustering and lower velocities, are closer to the linear approximation on the scales where the reconstruction is performed. For the \fofr{} dominated simulations, the opposite would then be the case. This is further highlighted for the pure \ISIS{} simulations, as shown in Fig.~\ref{fig:f4_recon_hist}, where the deviations from GR are larger. As a test, we also ran the reconstruction process for this simulation when varying the growth rate, $f$, by $10-20\%$ (as for Fig.~\ref{fig:f4_v_vary_f}). This did not lead to a significant difference in the histogram, but it did show up in the quadrupole as expected, and explained, by \cite{Nadathur2019recon}.

Figure~\ref{fig:ccf_theory_jackknife_xi2_recon} displays the quadrupole of the void-halo CCF for both the simulation data and the theory when the reconstruction step has been performed. In practice, this means that we performed the analysis as before, but instead of using the actual real space positions to calculate $\xi^r$, we instead used the reconstructed real space information. This mimics a possible procedure for performing the quadrupole analysis when using observational data. The model does, however, also require the density profile of the CDM, the mean outflow velocity profile, and the velocity dispersion. For this, we used the same as before, which was calculated from the actual real space data. This works as an approximation, as these quantities are only templates given to the model. For the velocity dispersion, it was also necessary to use the real space simulation data, as we did not have velocity information for the halos in reconstructed real space.

Comparing Figs.~\ref{fig:ccf_theory_jackknife_xi2} and \ref{fig:ccf_theory_jackknife_xi2_recon}, it is clear that the model better fits the data when the reconstruction step is not involved. This is expected, as reconstruction only approximates the real space halo positions, and can not fully recreate the real space CCF. The deviation between the model and data does, however, appear to be comparable for the different \ANUBISIS{} simulations.

\subsection{MCMC fits}
\begin{figure*}
    \centering
    \includegraphics[scale=0.783]{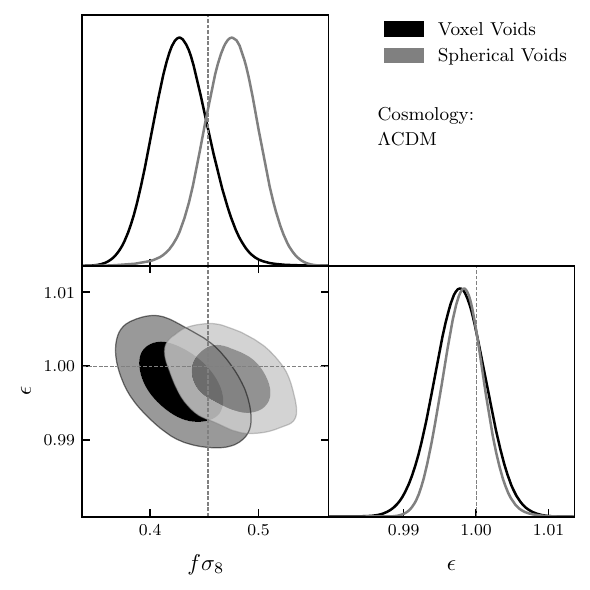}
    \includegraphics[scale=0.783]{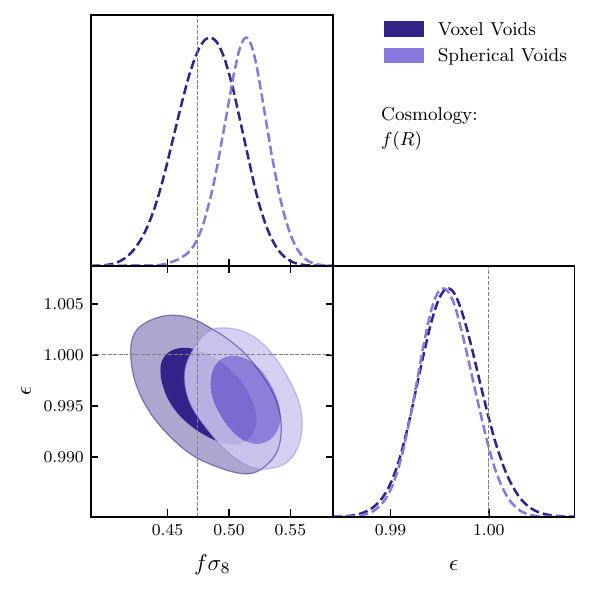}\\
    \includegraphics[scale=0.783]{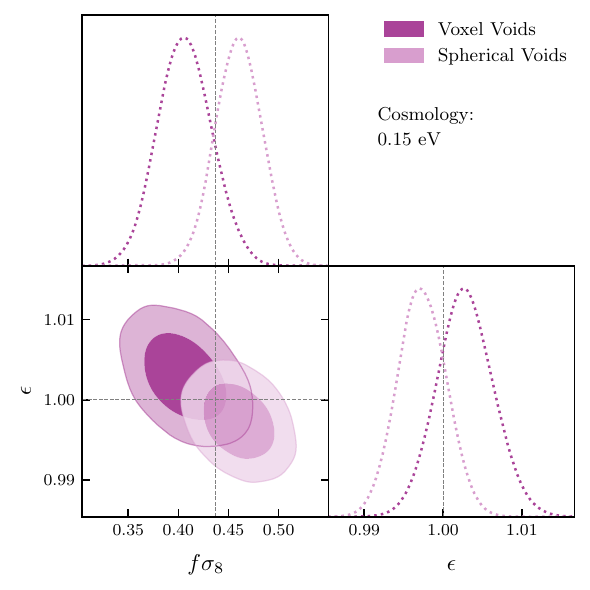}
    \includegraphics[scale=0.783]{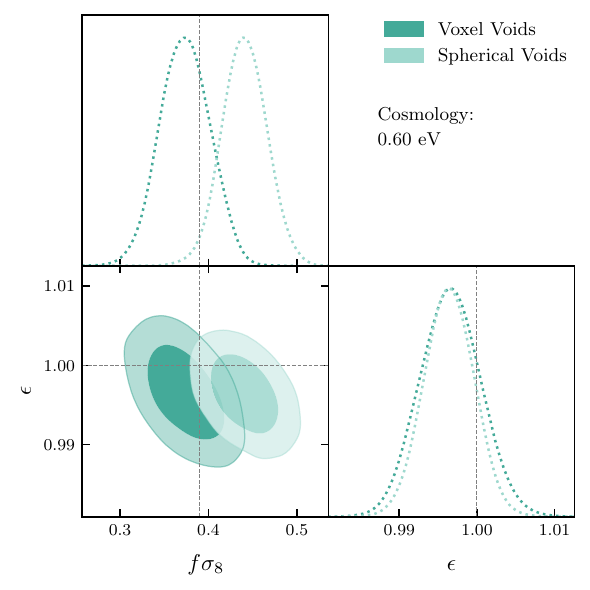}\\
    \includegraphics[scale=0.783]{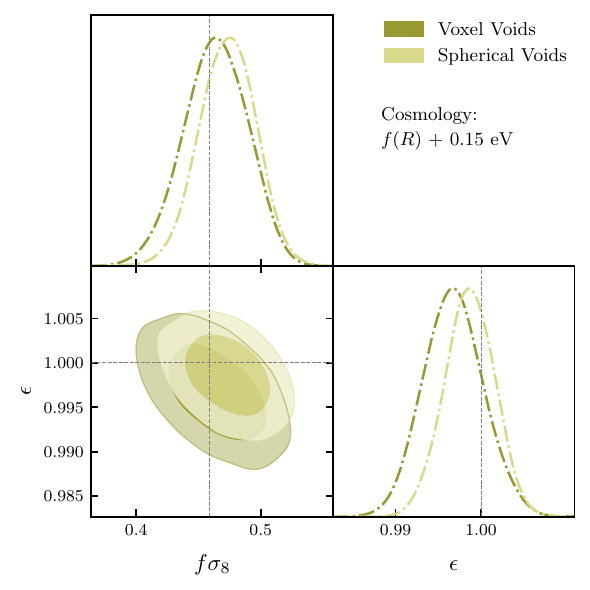}
    \includegraphics[scale=0.783]{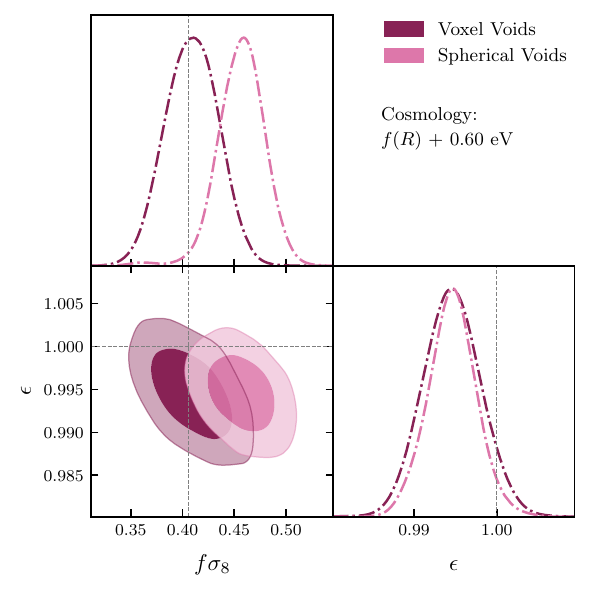}
    \caption{MCMC fits for the \ANUBISIS{} simulations. Each triangle plot shows the $f\sigma_8$ and $\epsilon$ estimates of one simulation for both the voxel and spherical void definitions. The grey dashed lines show the fiducial values and the contours display the $68\%$ and $95\%$ confidence intervals.} 
    \label{fig:mcmc}
\end{figure*}

A way to test the CCF theory for the different simulated cosmologies is to check if they all reproduce the original parameter values of the simulations in a similar manner. Figure~\ref{fig:mcmc} displays the posterior distributions of the $f\sigma_8$ and $\epsilon$ parameters based on the void-halo CCF simulation data and model calculations for all \ANUBISIS{} simulations. Each triangle plot shows the result for one simulation and both void definitions and follows the regular colour and linestyle pattern. The contours display the $68\%$ and $95\%$ confidence intervals and the grey dashed lines show the fiducial values, although with an assumed value of $f\approx\Omega_m^{0.55}$ for all simulations. We only display the $f\sigma_8$ and $\epsilon$ parameters, as the LOS velocity dispersion, $\sigma_{v_\parallel}$, is not an observable. The corresponding parameter values are listed in Table~\ref{table:mcmc_fits}.

It is evident from Fig.~\ref{fig:mcmc} that the voxel voids more closely reproduce the fiducial parameter values overall. This could possibly be because the voxel void definition results in a smoother CCF that is easier to fit than the very sharply varying spherical void CCF. We do, however, see that the model prediction is able to follow the sharp features (e.g. Fig.~\ref{fig:ccf_theory_jackknife_xi2}). The spherical void catalogue also contains a larger amount of voids with effective radii at the lower end of the included radius range, compared to the voxel void catalogue (Fig.~\ref{fig:nr_own_sims_hist}). It should be further explored whether or not performing the fits in different radius bins would alleviate the discrepancies between the spherical and voxel voids' abilities to reproduce the fiducial parameter values. This investigation is saved for future work. 

In general, we find that the spherical voids lead to higher $f\sigma_8$-values, compared to the voxel voids. This might be due to a large number of small voids with deep density profiles identified by the spherical void finder. This gives, on average, a higher velocity peak, as can be observed in Fig.~\ref{fig:vr_sim_theory} and ~\ref{fig:f4_vr_sim_theory}, and could lead to larger inferred values of $f\sigma_8$. The $\epsilon$-values, on the other hand, seem to be quite consistent between the two definitions for all simulations. The fact that the voxel and spherical voids coincide the best in their prediction for the \fofr$\,+\,0.15\,\mathrm{eV}$ simulation is most likely a coincidence, which illustrates how the different model components and void definitions can vary. In general, even though there is a difference between the void definitions, there does not seem to be a clear distinction for how well the fiducial values are reproduced for the various simulated cosmologies.

\begin{figure*}
    \centering
    \includegraphics[scale=0.783]{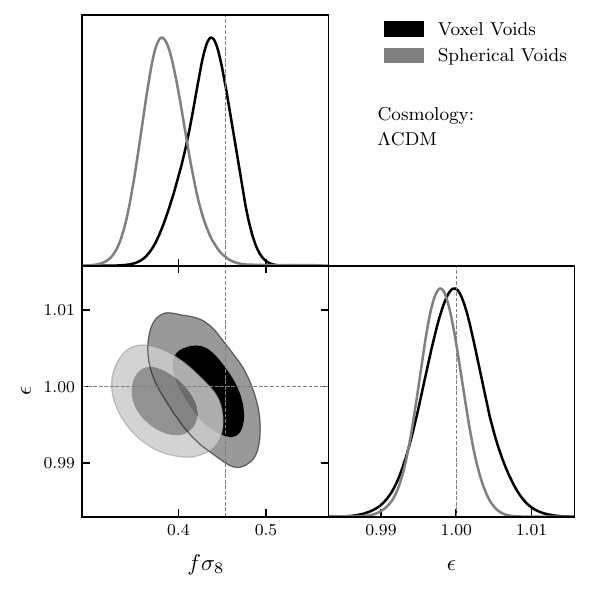}
    \includegraphics[scale=0.783]{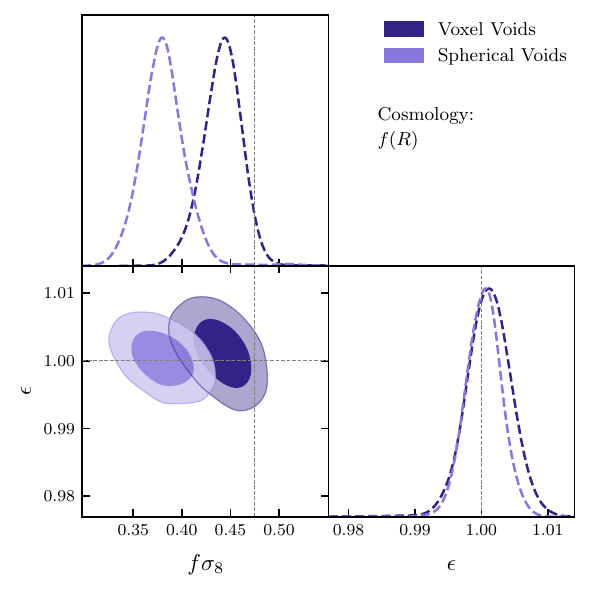}\\
    \includegraphics[scale=0.783]{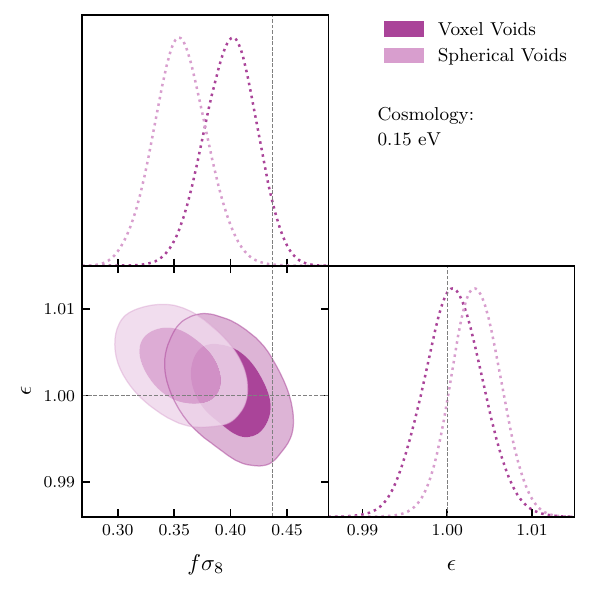}
    \includegraphics[scale=0.783]{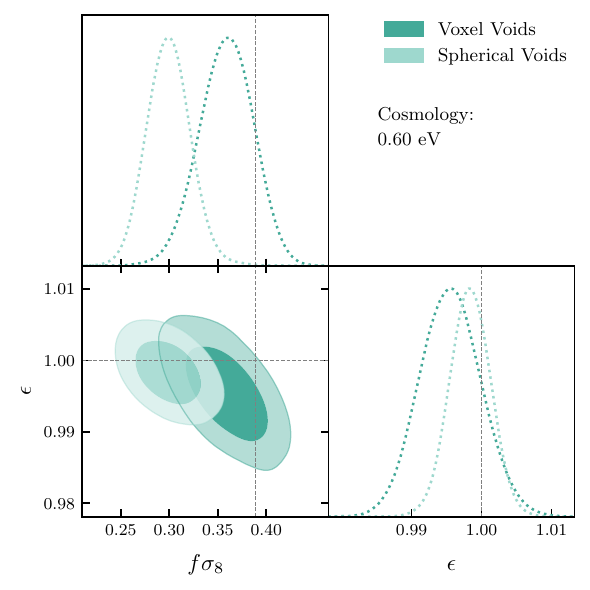}\\
    \includegraphics[scale=0.783]{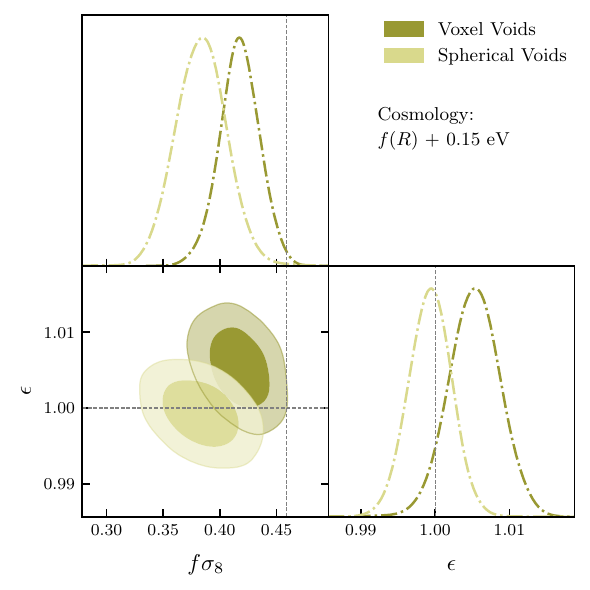}
    \includegraphics[scale=0.783]{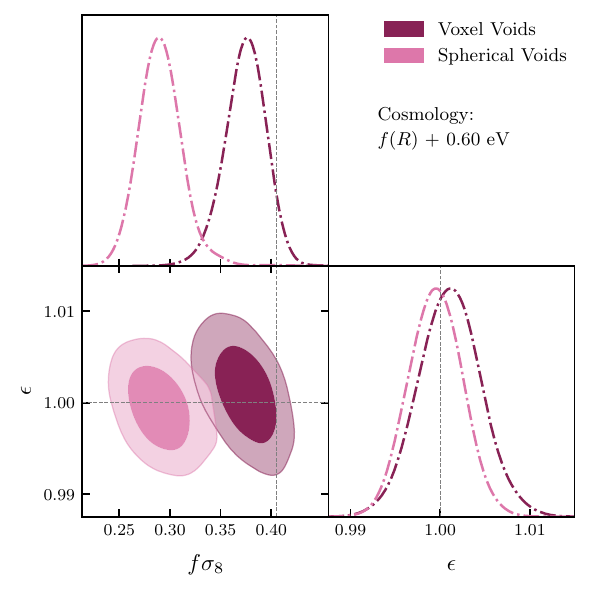}
    \caption{MCMC fits for the \ANUBISIS{} simulations with reconstruction. Each triangle plot shows the $f\sigma_8$ and $\epsilon$ estimates of one simulation for both the voxel and spherical void definitions. The grey dashed lines show the fiducial values and the contours display the $68\%$ and $95\%$ confidence intervals.} 
    \label{fig:mcmc_recon}
\end{figure*}

In Fig.~\ref{fig:mcmc_recon}, we see the results of the MCMC fits when the reconstruction step is included and $\beta$ is allowed to vary. The numerical values are given in Table~\ref{table:mcmc_fits}. Again, the voxel void definition better reproduces the fiducial values of the \ANUBISIS{} simulations. In fact, the differences between the voxel and spherical void definitions have increased. We still see quite consistent $\epsilon$-values between both void definitions, with a slightly lower value for the spherical voids except for the $0.15\,\mathrm{eV}$ and $0.6\,\mathrm{eV}$ cases. The $f\sigma_8$-values, however, are now consistently lower for the spherical voids, compared to the voxel voids, opposite of before. This might lie in the ability of the reconstruction method to correctly predict the high velocities found in the smaller spherical voids. Yet again, there is no clear difference between the simulated cosmologies and how accurately the fiducial parameter values are recovered. This underlines that the reconstruction method should be further studied for various void definitions.

It should be noted that for the fits that include reconstruction and a free $\beta$-value, we encounter the issue of a multimodal likelihood surface for $\beta$. This issue is more prominent for the voxel void definition than for the spherical one. It does, however, seem to be an artefact of the reconstruction model and, as a first approach, we dealt with it by constraining the range of the flat prior. The new range was chosen to include the most likely values resulting from the parameter fit performed with the original prior.

\begin{table*}
\caption{Best fit value for $f\sigma_8$ and $\epsilon$ for all \ANUBISIS{} simulations for both the voxel (VV) and spherical (SV) void definition, with and without reconstruction. The limits are given at $68\%$ and the flat prior ranges are $f\sigma_8\in[0.08,1.3]$ and $\epsilon\in[0.8,1.2]$.}             
\label{table:mcmc_fits}      
\centering                          
\begin{tabular}{c c c c c c c c c}        
\hline\hline  
\\[-0.32cm]
Simulation &$f\sigma_{8,VV}$ &$f\sigma_{8,SV}$ & $\epsilon_{VV}$& $\epsilon_{SV}$ & $f\sigma_{8,VV}^\mathrm{recon}$ & $f\sigma_{8,SV}^\mathrm{recon}$  & $\epsilon^\mathrm{recon}_{VV}$ & $\epsilon^\mathrm{recon}_{SV}$\\[0.05cm]
\hline
\\[-0.3cm]
    \vspace{+0.15cm}
   \texttt{lcdm} & $0.43^{+0.03}_{-0.03}$ & $0.47^{+0.03}_{-0.03}$&$0.998\pm 0.004$ & $0.998\pm 0.004$& $0.43^{+0.03}_{-0.03}$& $0.39^{+0.03}_{-0.03}$& $1.000\pm 0.005$&$0.999\pm 0.003$\\
   \vspace{+0.15cm}
   \texttt{fofr} & $0.48^{+0.03}_{-0.03}$&$0.51^{+0.02}_{-0.02}$& $0.996\pm 0.004$&$0.996\pm 0.003$ &$0.44^{+0.03}_{-0.02}$&$0.38^{+0.02}_{-0.03}$& $1.001\pm 0.004$&$1.000\pm 0.003$\\
   \vspace{+0.15cm}
   \texttt{015ev}&$0.41^{+0.03}_{-0.03}$ &$0.46^{+0.03}_{-0.03}$& $1.003\pm 0.004$& $0.997\pm 0.004$&$0.40^{+0.03}_{-0.03}$& $0.36^{+0.03}_{-0.03}$& $1.001\pm 0.004$&$1.003\pm 0.003$\\
   \vspace{+0.15cm}
   \texttt{06ev} &$0.37^{+0.03}_{-0.03}$ &$0.44^{+0.03}_{-0.03}$& $0.997\pm 0.004$&$0.996\pm 0.004$ &$0.36^{+0.03}_{-0.03}$ & $0.30^{+0.03}_{-0.03}$& $0.996\pm 0.005$&$0.998\pm 0.003$\\
   \vspace{+0.15cm}
   \texttt{fofr\_015ev} & $0.46^{+0.03}_{-0.03}$& $0.47^{+0.03}_{-0.03}$& $0.997\pm 0.004$&$0.999\pm 0.003$& $0.42^{+0.02}_{-0.02}$& $0.38^{+0.03}_{-0.03}$& $1.005\pm 0.004$&$0.999\pm 0.004$\\
   \vspace{+0.1cm}
   \texttt{fofr\_06ev}  & $0.41^{+0.03}_{-0.03}$& $0.46^{+0.03}_{-0.02}$& $0.995\pm 0.004$&$0.995\pm 0.004$& $0.37^{+0.03}_{-0.02}$& $0.29^{+0.02}_{-0.03}$& $1.001\pm 0.004$&$1.000\pm 0.003$\\
\hline                                   
\end{tabular}
\end{table*}

\section{Conclusions}
\label{sec:conclusions}

Cosmic voids are promising and independent probes of gravity and cosmology expected to provide stringent constraints on $f\sigma_8$ and the ratio between the Hubble distance and transverse comoving distance in upcoming space missions such as \euclid{} \citep{Hamaus2022,Radinovic2023}. Voids are also proposed as grounds for studying the effects of massive neutrinos and modified gravity, both separately and simultaneously, and as a way to break known degeneracies \citep{Li2012,Massara2015,Voivodic2017,Kreisch2019,Perico2019,Contarini2021}. In this paper, we investigate the performance of models describing the void-halo CCF, the radial mean void velocity profile, and a reconstruction method using \ANUBISIS{} simulations with both massive neutrinos and Hu-Sawicki \fofr{} gravity. Our simulation suite consists of six simulations, one reference \LCDM{} cosmology, one with Hu-Sawicki \fofr{}-modified gravity ($|f_{R0}|=10^{-5}$), two with massive neutrinos ($\sum m_\nu = 0.15\,\mathrm{eV},\, 0.6\,\mathrm{eV}$), and two with both massive neutrinos and modified gravity ($|f_{R0}|=10^{-5}$ + $\sum m_\nu = 0.15\,\mathrm{eV},\, 0.6\,\mathrm{eV}$). Occasionally, we supplement with a \LCDM{} and \fofr{} simulation ($|f_{R0}|=10^{-4}$) run with \ISIS{}, when we want to investigate the effects of modified gravity further.
 
We find that the linear void velocity model (Eq.~\ref{eq:vr_lin}) fits the velocity profiles calculated directly from the data similarly for all \ANUBISIS{} simulations. This is the case both when using the voxel void and spherical void definitions to locate voids in the halo catalogues. In fact, the increased velocities in the \fofr{} dominated simulations happen to partially compensate for the bias between the velocity profile calculated from the data and the one calculated from linear theory, arising from a sparse tracer sample \citep{Massara2022}. This shows that the linear velocity model needs to be improved further before differences due to massive neutrinos or modified gravity can be seen clearly. We also fit the velocity profile from the simulation data to a more general velocity model (Eq.~\ref{eq:general_vr}), with one free parameter. This model fits the data well and has a different best-fit parameter value for the various simulations. If the radial velocity profile could be observed directly with good precision from galaxy surveys, the value of this free parameter would be an interesting observable. The value is, however, degenerate for massive neutrinos and \fofr{}-modified gravity. Similarly, we find that the LOS velocity dispersion profiles have the same shape for all the simulated cosmologies, but different amplitudes. This amplitude is an interesting parameter, but it has the same issue of degeneracy, in addition to not being readily observable. 

We compare the monopole, quadrupole, and hexadecapole in redshift space calculated from the simulation data with two theoretical model outcomes calculated by \victor{}. The only difference in the two models is the velocity profile input, which in one case is the radial linear velocity profile and in the other case, the velocity profile calculated from the simulation data. For all the multipoles, we see that the model fits the data with similar accuracy for all the different simulations. As expected, there is a preference towards the model that has the velocity profile from the data given as input, both for the voxel and spherical void definitions. As it stands, we need more accurate modelling and high-resolution surveys if we hope to use the void-galaxy CCF to put constraints on \fofr{}-modified gravity or massive neutrinos. However, our simulations show that modified gravity and massive neutrinos affect the monopole and quadrupole in opposite ways, making the void-galaxy CCF yet another degenerate observable. 

When calculating the model prediction of the void-halo CCF in redshift space, we have in the above case used the real space CCF information from the simulation boxes as model input. For data obtained from a galaxy survey, this information is unknown and can be approximated through a reconstruction process where the redshift space positions are put back into real space by solving for the displacement field \citep{Nadathur2019recon}. When performing reconstruction for all simulations, we find that the process performs similarly for all simulated cosmologies, although slightly better for the most massive neutrino case and slightly worse for the modified gravity case, compared to \LCDM{}. This could be due to the reduced clustering and lower velocities in the massive neutrino simulations better adhering to the linear approximations of the reconstruction model, and the opposite for the modified gravity simulation. When fitting the CCF model to the data when the reconstruction step is included, the model provides a slightly worse fit to the data. This is expected as we are now providing the model with information that only approximates the actual real space CCF that is needed as model input. The model-data fit is, however, comparable for all the \ANUBISIS{} simulations also in this scenario. 

During our analysis, we kept the growth rate, $f$, constant in all cases, effectively ignoring the scale dependence for modified gravity and massive neutrino cosmologies. We used the expected \LCDM{} value, $f=\Omega_m^{0.55}$, for all simulations to keep the study consistent, and instead investigated if this choice gives visible biases in our model fits. As recapped above, all model-data fits perform similarly for the different simulations. The linear velocity profile does, however, show a slight improvement for the \fofr{} simulations, especially visible for the pure \ISIS{} run. We, therefore, tried varying the value of $f$ in this model to see if this could further improve the fit. For the voxel void definition, which already overestimates the clustering, increasing $f$ to values more in line with expectations for \fofr{}-modified gravity, only contributes to further mismatch between the model and the simulation data. For the spherical void definition, which suffers less from the effect of a sparse tracer sample, increasing $f$ can give a better fit. This only shows that the choice of void finding algorithm and the current sparse tracer sample issue in the void velocity modelling is important and must be consistently handled before possible effects of a different growth rate in the various simulated cosmologies can be resolved. 

For all the \ANUBISIS{} simulations, we performed MCMC fits for $f\sigma_8$ and the Alcock-Paczy\`{n}ski parameter, $\epsilon$. We did this both with and without reconstruction. In both cases, we find that the voxel void definition better recovers the fiducial values of the $f\sigma_8$-parameter for the various simulations. The $\epsilon$-values are more consistently recovered between the two void definitions. The discrepancies are most prominent when reconstruction is included, where the $f\sigma_8$-values recovered for the spherical voids are consequently underestimated for all simulated cosmologies, as opposed to slightly overestimated without the reconstruction step. Still, for both void definitions, the ability to recover the fiducial parameter values is similar for the different simulations. There is no indication that the parameter estimations are more accurate for a \LCDM{} cosmology, compared to \fofr{}-modified gravity or massive neutrinos. 

Our investigations suggest that the current limitations of the void theories must be dealt with in order for the models to be accurate enough to clearly showcase differences between the various simulated cosmologies. This is particularly clear for the velocity profile modelling and the reconstruction method. In addition, higher resolution simulations are necessary to lower the uncertainties in the data. Once this is in place, cosmic voids are a promising ground for studying cosmologies with massive neutrinos and \fofr{}-modified gravity.

\begin{acknowledgements}
      We thank the Research Council of Norway for their support. Parts of our computations were performed on resources provided by UNINETT Sigma2 -- the National Infrastructure for High Performance Computing and Data Storage in Norway. We also wish to thank Sla{\dj}ana Radinović for her contributions  in the form of valuable input, discussions and instructions regarding the void component of this paper.
\end{acknowledgements}

\bibliographystyle{aa} 
\bibliography{main} 

\begin{appendix}

\section{ANUBISIS resolution}
\label{sec:anubisis_res}
\begin{figure}
    \centering
    \includegraphics{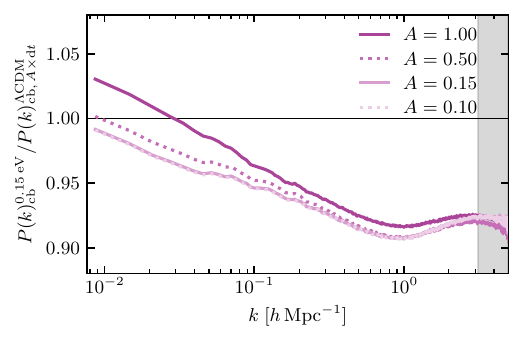}\\
    \caption{Matter power spectrum ratio between a $0.15\,\mathrm{eV}$ massive neutrinos simulation and a \LCDM{} simulation with varying timestep, performed by \ANUBIS. The simulation box has $L_\mathrm{box} = 1024\,h^{-1}\mathrm{Mpc}$, $N_\mathrm{cdm} = N_\nu = 1024^3$, and a fixed coarse base grid of $512^3$. The time step is calculated by the \RAMSES{} algorithm for both runs, but for the \LCDM{} case it is multiplied by a reduction factor, $A$. We see here that one way to compensate for the coarse base grid, especially at large scales, is to shorten the time step. This requires a longer run time for the \LCDM{} simulation, but not more memory as would be the result of adopting a finer grid.}
    \label{fig:timesteps}
\end{figure}

In the upper panel of Fig.~\ref{fig:pk_HMF_ratio_own_sims}, we present the matter power spectrum ratio for the non-\LCDM{} simulations, compared to the reference \LCDM{} run. We mention that at larger scales there is originally an excess in the power spectrum for the massive neutrino and \fofr{} + massive neutrino runs. This is attributed to the coarse base grid settings used in the simulations. With the resources available to us at the time of running \ANUBISIS{}, we made appropriate adjustments in order to constrain our memory and time usage. To accomplish this, we chose a base grid of $512^3$ for all the simulations, which results in a lower resolution at early times and at large scales where there is less clustering. This is due to the nature of the \RAMSES{} AMR-scheme, which creates more refinements as structure grows. In a ratio comparison, such an internal consequence of the algorithm should cancel out, but as the massive neutrinos contribute when calculating the time steps, the simulations containing these particles automatically operate with smaller time steps. This leads to a more detailed time development of the CDM particles in the massive neutrino and \fofr{} + massive neutrino simulations, compared to the \LCDM{} and pure \fofr{} runs. Effectively, we resolve the simulations containing massive neutrinos better at large scales, resulting in the refinement of more clustering in these cases. This evens out as time passes and more refinements are made, but for large neutrino masses, this also happens for small scales, where the suppressed structure formation leads to fewer refinements within the code, compared to the \LCDM{} case. To help even out the issues at large scales, we ran the \LCDM{} and pure \fofr{} simulations with one-fourth of the time step originally calculated by the code. The effect of adjusting the time step is illustrated in Fig.~\ref{fig:timesteps}. Ideally, the simulations should be run with a coarser base grid and with a higher particle density, which would also contribute to better resolution overall. This is not done in our case, due to computational costs. 

In the end, this issue is practically equivalent to having a slightly different $\sigma_8$-value ($2-3\%$ lower) at large scales. We have not accounted for this in our analysis, as the resulting changes are small. For our MCMC fits, the true $\sigma_8$-value is still encompassed in the flat prior interval, and the only difference in Figs.~\ref{fig:mcmc} and \ref{fig:mcmc_recon} would be a slight adjustment of the fiducial $f\sigma_8$-line towards lower values for \LCDM{} and \fofr{}. Simulations with the same issue still showed good agreement with other codes in the MNCCP \citep{MNCCP}.

\section{Quadrupole shape}
\label{sec:quad_shape}
If the redshift space void-halo CCF model (Eq.~\ref{eq:streaming}) did not include a velocity dispersion for the mean outflow velocity profile, the shape of the quadrupole in redshift space would be qualitatively similar for all void definitions. This would mean a trough inside the void due to outflow towards the overdensity at the edge, and a peak outside due to infall towards the same overdensity, akin to the shape we see for the voxel voids in Fig.~\ref{fig:ccf_jackknife}. Adding a velocity dispersion to the model effectively corresponds to `smoothing' the RSD quadrupole.

\begin{figure}
\hspace{-0.4cm}
    \centering
    \includegraphics{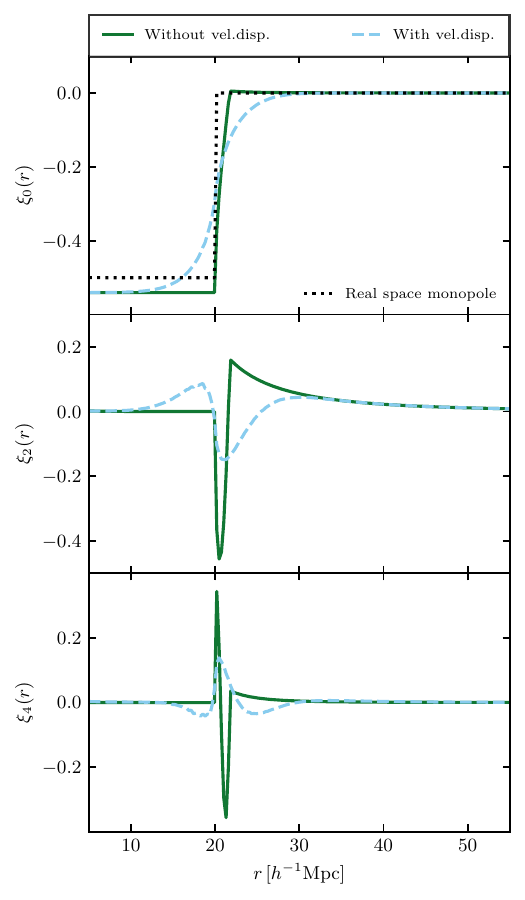}
    \caption{Monopole, quadrupole, and hexadecapole for the toy model of the void density profile both with (blue dashed line) and without (green line) velocity dispersion. The flat density profile inside the void results in a sharp feature in the quadrupole at the void radius, which smoothes out and gives a peak right before the trough inside the void.}
    \label{fig:ccf_shape}
\end{figure}

If we take the real space radial coordinate, $r$, to be equal to the redshift space coordinate, $s$, in the theory model, then from Eq.~(\ref{eq:streaming_coord_change}) it can be shown that for $v_r/(raH) \ll 1$ we roughly have $\xi^s_2 \propto \frac{d}{dr}(v_r/raH)$. This means that it is deviations from linearity of the void velocity profile (equivalent to deviations from a flat density profile) that lead to a quadrupole inside the void. To better understand the shape of the quadrupole for different void definitions, it is useful to consider a simple toy model. We define a density profile that is perfectly flat inside the void, and equal to zero outside: $\delta=\delta_0$ for $r<R_\mathrm{void}$ and $\delta=0$ for $r>R_\mathrm{void}$. This then gives a velocity profile $v_r\propto r$ inside the void, and $v_r\propto R_\mathrm{void}^3/r^2$ outside, if we apply Eq.~(\ref{eq:vr_lin}). This again gives $\xi_2^s=0$ inside the void, and $\xi_2^s\propto R_\mathrm{void}^3/r^4$ outside. At the void edge, we have a sharp feature. For real voids, the density profile is never perfectly flat on the inside, and the deviations break the linearity of the outflow velocity, resulting in a trough in the quadrupole inside the void. The flatter the density profile is inside the void, the sharper the peak of the velocity profile is, leading to sharper features in the quadrupole. The monopole, quadrupole, and hexadecapole, in the absence of a velocity dispersion, are illustrated by the green line in Fig.~\ref{fig:ccf_shape}. The sharp features at $R_\mathrm{void}$ are visible both for the quadrupole and hexadecapole.

The above scenario is the case when there is no velocity dispersion. If we add this to the model, the imposed smoothing effect simply reduces the amplitude of the no-dispersion quadrupole for the voxel void definition. This means that the smoothing length given by the size of the velocity dispersion is smaller than the features seen in the no-dispersion quadrupole. On the other hand, for spherical voids, where the density profile is very flat in the void centre and the quadrupole has sharp features as a consequence, this smoothing results in a sharp peak followed by a sharp trough close to the void edge (as seen in Fig.~\ref{fig:ccf_jackknife}). This shows up at the point where the density profile stops being flat, which is also the point where the velocity profile has a sharp peak. The blue dashed line in Fig.~\ref{fig:ccf_shape} shows this effect clearly for the quadrupole. The smaller the average void is, the larger this first peak is, as the smoothing effect is more dramatic. This shape is also similar to what previous studies have found for Zobov voids \citep{Nadathur2019recon,Nadathur2019,NadathurPercival2019,Massara2022,Woodfinden2022}, which is not surprising, as these voids also have a fairly flat density profile, $\delta\sim-1.0$, at the void centre.

\end{appendix}

\end{document}